\documentclass{aa}  

\usepackage{graphicx}
\usepackage{mathtools}
\usepackage{amsmath}
\usepackage{txfonts}
\usepackage[table, dvipsnames]{xcolor}
\usepackage{colortbl}
\usepackage[normalem]{ulem}
\def\msun{{ ~M}_{\odot}}
\def\rsun{{ ~R}_{\odot}}

\def\lsun{{ ~L}_{\odot}}

\def\gpy{{\rm ~Gpc}^{-3} {\rm ~yr}^{-1}}
\def\kms{{\rm ~km} {\rm ~s}^{-1}}

\begin{document} 

   \title{The impact of common envelope development criteria on the formation of LIGO/Virgo sources.}
   \author{A. Olejak
          \inst{1}
          K. Belczynski\inst{1}
          N. Ivanova \inst{2}
          }

   \institute{
            Nicolaus Copernicus Astronomical Center, Polish Academy of Sciences, ul. Bartycka 18, 00-716 Warsaw, Poland\\
            \email{aolejak@camk.edu.pl,chrisbelczynski@gmail.com}
         \and
             Department of Physics, University of Alberta, Edmonton, AB, T6G 2E7, Canada
             }

   \date{Received ; accepted }

 
  \abstract
{
The treatment and criteria for development of unstable Roche lobe overflow 
(RLOF) that leads to the common envelope (CE) phase have hindered the evolutionary
predictions for decades. In particular, the formation of black hole--black 
hole (BH-BH), black hole--neutron star (BH-NS), and neutron star--neutron star
(NS-NS) merging binaries depends sensitively on the CE phase in classical 
isolated binary evolution model. All these mergers are now reported as LIGO/Virgo
sources or source candidates. CE is even considered by some as a mandatory 
phase in the formation of BH-BH, BH-NS or NS-NS mergers in binary evolution models. 
At the moment, there is no full first-principles model for development 
of CE. 
We employ the {\tt StarTrack} population synthesis code to test the current 
advancements in studies on stability of RLOF for massive donors to assess
their effect on LIGO/Virgo source population. In particular, we allow for
more restrictive CE development criteria for massive donors ($M>18\msun$). 
We also test a modified condition for switching between different types
of stable mass transfer, thermal or nuclear timescale. 
Implemented modifications significantly influence basic properties of 
merging double compact objects, sometimes in non-intuitive way. For one of 
tested models with restricted CE development criteria local 
merger rate density for BH-BH systems increased by a factor of 2-3 due to 
emergence of a new dominant formation scenario without any CE phase. 
We find that the changes in highly uncertain assumptions on RLOF physics may 
significantly affect (i) local merger rate density, (ii) shape of the mass
and mass ratio distributions, and (iii) dominant evolutionary formation 
(with and without CE) scenarios of LIGO/Virgo sources. Our results demonstrate 
that without sufficiently strong constraints on RLOF physics, one is
not able to draw fully reliable conclusions about the population of double 
compact object systems based on population synthesis studies.
}


  \maketitle
%

\section{Introduction}

The concept of close binary systems' formation, such as merging double compact
objects (DCO) or X-ray binaries, through the common envelope (CE) phase, begun
to appear in the literature almost 50 years ago 
\citep{1976IAUS...73...75P,1976IAUS...73...35V}. In this concept, one binary 
component enters the other binary companion's envelope. The orbital energy is 
transferred to the envelope due to various drag forces, resulting in the binary 
orbit shrinking.  Finally, the envelope either can be ejected from the system, 
leaving behind a newly formed close binary system, or the two stars merge. 
One of the approaches to find the outcome of a CE event without doing detailed 
3D simulations is by considering the energy budget of a CE event and hence is 
known as energy formalism. In its parameterized form, it has been introduced by 
\cite{1984ApJ...277..355W} and \cite{1988ApJ...329..764L}, and is known 
as ``$\alpha_{\rm CE}$-formalism''. Due to its simplicity, $\alpha_{\rm CE}$-
formalism is still widely used in population synthesis studies 
\citep[e.g.,][]{alej2020commonenvelope}. This simplified form of energy formalism 
equates only two energies: the envelope's binding energy (the energy required to 
eject the envelope to infinity) and the change in the orbital energy (the available 
energy source). This allows to estimate the orbital separation after a CE event, 
assuming that envelope has been ejected:
    
    \begin{equation} 
    \frac{G M_{\rm don,i}  M_{\rm don,env}}{\lambda R_{\rm don,lob}} = 
    \alpha_{\rm CE}  \left(\frac{G M_{\rm don,f} M_{\rm comp,f}}{2 a_{\rm f}}
    -\frac{G M_{\rm don,i} M_{\rm comp,i}}{2 a_{\rm i}} \right),  \label{eq:CE}
    \end{equation}
   
\noindent Here $a_{\rm{i/f}}$ -- initial or final orbital separation, $M_{\rm don,i/f}$ -- initial 
or final donor mass, $M_{\rm comp,i/f}$ -- initial or final companion mass respectively,
$G$ -- gravitational constant, $M_{\rm don,env}$ -- mass of the donor envelope, $R_{\rm don,lob}$ -- 
Roche lobe radius of the donor at the onset of Roche-lobe Overflow (RLOF),
i and f - initial and final values of mass and separation, and $\lambda$ - is a
measure of the donor central concentration 
\citep{1990ApJ...358..189D,2000A&A...360.1043D,Xu_2010}. 

Equation \ref{eq:CE}  postulates that the transfer of binary system orbital energy into 
the energy of the envelope takes place with some efficiency
$\alpha_{\rm CE}$, which, unless other energy sources are present, can not be more than one.
Simulations and observations of CE phase indicates that the value of parameter is typically 
$\alpha_{\rm CE}<0.6-1.0$ \citep{Zuo_2014,2016MNRAS.460.3992N,2017MNRAS.470.1788C,Iaconi_2019}. 
On the other hand, if the simulations of a CE event are performed while including more physical 
processes, for example, accretion, the effective value of $\alpha_{\rm CE}$ can be as high as 
5 \citep[e.g., see][]{Fragos_2019}. A similar effect of increasing the apparent CE efficiency 
to more than one be produced by exotic nucleosynthesis and jets 
\citep{2010MNRAS.406..840P,10.1093/mnras/stz2013,zevin2020channel,grichener2021common}.
Some population synthesis studies have already adopted such high values
\citep[e.g., see][]{Santoliquido_2021}.

The CE efficiency parameter $\alpha_{\rm CE}$ and the binding energy parameter $\lambda$ 
are often coupled in population synthesis studies or in observations' analysis when energy 
formalism is applied. At the same time, each of them is subject to many uncertainties.
For example, $\lambda$  was introduced to relate the ``true'' binding energy of the 
envelope (as obtainable from detailed stellar models) to its simple parameterized form. 
However, what is the true binding energy of the donor, is still the subject of discussion 
\citep[e.g., see \S3 in][]{comenv_book_2020}.
The CE efficiency parameter is expected to depend on the system's specifics and which 
physical processes of energy creation or energy loss took place (e.g., on the system's 
mass ratio, the evolutionary stage of the donor, or the companion's nature). Derived from 
observations, values of $\alpha_{\rm CE}$ seem to be systematically lower for asymptotic 
giant branch (AGB) donors than for red giant branch (RGB) donors \citep{Iaconi_2019}, 
which have different internal structure and envelope binding energy. AGB donors 
envelopes are considered to be less tightly bound to the core than those of a RGB donors 
\citep{1968AcA....18..255P,1994MNRAS.270..121H}. However, even taking into account this 
AGB stars feature does not sufficiently help in successive envelope ejection 
\citep{2020A&A...644A..60S}. Numerical simulations encounter difficulties in successful 
envelope ejection unless some other energetic process (except orbital energy release) 
is included \citep[e.g.,][]{2012ApJ...744...52P,2017MNRAS.464.4028I,2020A&A...644A..60S}. 
Also recent studies of \cite{Klencki2020} confirmed that even with the most favorable 
assumptions, a successful CE ejection in BH binaries is only possible if the donor is a 
massive convective-envelope giant. On the other hand, massive stars (BH progenitors) 
may be a subject of extensive mass loss through enhanced winds before they reach the 
RSG stage what may even cause the spontaneous envelope loss 
\citep[e.g.,][]{1991A&A...252..159V,1998NewA....3..443V,2002ApJ...575.1037E}.

The energy formalism in its parameterized form is a convenient way to predict CE outcomes, 
but, as argued now, it is not necessarily a well-founded method  \citep{comenv_book_2020}. 
While at the moment, there is no complete understanding of the CE evolution in all the cases, 
it is understood that the CE event could be preceded by stable mass transfer (MT) on 
different timescales.
The criteria for the occurrence of CE phase are still under development as well.

New stellar mass loss models \citep{Ge2010,Ge2015,Ge2020a,Ge2020b} and detailed simulations for close, mass exchanging binaries \citep{2017MNRAS.465.2092P,Misra2020} 
have shown that RLOF may be stable over a much wider parameter space than previously thought. 
The same studies indicate that RLOF stability depends not only on the system mass ratio and the 
envelope type (convective or radiative), but also e.g., on the metallicity, stellar type or 
radius of the donor star. Those results begin to find confirmations when comparing theoretical 
models with the observed systems \citep{Cherepashchuk_2019,leiner2021census}. The summary of 
the recent progress on RLOF stability has been summarized in \S 2.2 of \cite{Klencki2020}.
Unfortunately, due to scarcity of observations for massive stellar systems during the ongoing 
RLOF phase and the high calculation costs, the current CE study (observations and simulations) 
usually refers to low-mass binary systems \citep{2015MNRAS.450L..39N, 2016MNRAS.460.3992N,2020arXiv200103337J}. 
Such systems are not progenitors of BH-BH, BH-NS or NS-NS binaries.
   
CE phase is a key element setting the formation of DCOs in the classical isolated 
binary evolution channel and therefore understanding of CE is crucial in studies 
of origin of merging DCOs. Recently more and more signals form BH-BH, BH-NS and
NS-NS mergers have been detected by LIGO/Virgo instruments 
\citep{Abbott_2019,catalog03} during the O1, O2 and O3 runs with reported 
parameters of systems such as the masses, spins and redshifts. It is still 
unknown what fraction of gravitational wave (GW) signal mergers formed through isolated
binary evolution in the field ~\citep{Bond1984b,Tutukov1993,Lipunov1997,
Voss2003,Belczynski2010a,Dominik2012,Kinugawa2014,2014A&A...564A.134M,
Hartwig2016,Spera2016,Belczynski2016b,
Eldridge2016,Woosley2016,Stevenson2017,Kruckow2018,Hainich2018,Marchant2018,Spera2019,
Bavera2020}, the dense stellar system dynamical channel~\citep{Miller2002a,
Miller2002b,PortegiesZwart2004,Gultekin2004,Gultekin2006,OLeary2007,Sadowski2008,
Downing2010,Antonini2012,Benacquista2013,Bae2014,Chatterjee2016,Mapelli2016,Hurley2016,
Rodriguez2016a,VanLandingham2016,Askar2017,Morawski2018,
Banerjee2018,DiCarlo2019,Zevin2019,ArcaSedda2017,Rodriguez2018a,Perna2019,Kremer2020}; 
isolated multiple (triple, quadruple) systems  
~\citep{Antonini2017b,Silsbee2017,Arca-Sedda2018,LiuLai2018,
Fragione19}, mergers of binaries in galactic nuclei \citep{Antonini2012b,Hamers2018,
Hoang2018,Fragione2019b}; the chemically homogeneous evolution channel consisting of 
rapidly spinning stars in isolated binaries ~\citep{deMink2016,Mandel2016a,Marchant2016,
Buisson2020} or Population III origin DCO binary mergers 
\citep{Bond1984b,Kinugawa2014,Tanikawa:2020cca}. 

Over the years different groups developed their population synthesis codes and 
try to put better constrains on astrophysical processes by comparing theoretical 
model results with the known Galactic and extragalactic compact object population 
\citep[e.g.,][]{1996A&A...309..179P,Hurley_2002,Kruckow2018,2020A&A...636A.104B,2020RAA....20..161H}. 
Several commonly encountered uncertainties in population synthesis studies strongly 
influence formation of DCOs and therefore, are subjects of active research. Examples of 
uncertain processes and parameters which are crucial for DCO mergers evolution are 
metallicity-specific star formation rate density 
\citep{2020A&A...636A..10C,Santoliquido2021,2021arXiv210302608B}, NS and BH natal kicks 
\citep{mandel2021} or MT during stable/unstable RLOF \citep{Vinciguerra2020,Howitt2020,Bavera2020}.

In this paper, we study how applying the most recent developments in the understanding 
of RLOF stability into the population synthesis affects DCOs formation. In \S 2 we describe 
the general method and the input physics implemented in the current version of {\tt StarTrack} 
population synthesis code. In \S 3 we introduce the revised CE development criteria and 
modified stable RLOF treatment examined in this paper. In \S 4 we present the results 
of our simulations: DCO local merger rate density, BH-BH and BH-NS mass ratio distributions 
and BH-BH mass distributions for three tested models.
\S 5 is a description of evolutionary scenarios leading to the BH-BH and BH-NS 
mergers formation in three tested models. It also includes three diagrams with an example of
systems evolution. \S 6 contains a brief discussion of our results together with 
the conclusions.

\section{Method} \label{sec: method}

To simulate formation and mergers of DCO systems in the local Universe we used 
updated {\tt StarTrack} population synthesis code, developed over the years 
\citep{2002ApJ...572..407B,2008ApJS..174..223B}.
The code allows to simulate evolution of a single star as well as binary star 
system for a wide range of initial conditions and physical parameters. Currently 
used version of implemented physics, adopted star formation history (SFH) and 
metallicity of the Universe is described in \cite{2020A&A...636A.104B} with two 
recent modifications, both explained in \S 2 of \cite{olejak2020origin}.
We adopted 3-broken power-law initial mass function (IMF) 
\cite{1993MNRAS.262..545K,2002Sci...295...82K}, weak pulsation pair-instability 
supernovae (PPSN) and pair-instability supernovae (PSN) 
\citep{2017ApJ...836..244W,2016A&A...594A..97B}.
We apply procedures for accretion onto a compact object during stable RLOF and 
from stellar winds based on the analytic approximations described in \cite{King2001} 
and \cite{Mondal2020}. For non-degenerate accretors we adopt 50\% non-conservative 
RLOF \citep{Meurus1989,Vinciguerra2020} with a fraction of the lost donor mass accreted onto 
the companion ($f_a=0.5$), and the rest of mass ($1-f_a$) leaving the system together with part of the
donor and orbital angular momentum \citep[see \S 3.4 of][]{2008ApJS..174..223B}.
We use 5\% Bondi-Hoyle rate accretion onto the compact object during the CE phase 
\citep{Ricker&Taam, MacLeod2015a,MacLeod2017}. The procedure is based on equations 5.3-5.7 of \citep{Bethe1998} and equations A1-A10 of \cite{2002ApJ...572..407B}, and has been recently summarized in Appendix B of \cite{olejak2020origin}.  
For stellar winds we use formulas based on theoretical predictions of radiation driven 
mass loss \citep{Vink2001} with inclusion of Luminous Blue Variable mass loss 
\citep{Belczynski2010a}.
All tested models have our standard physical values for the envelope ejection 
efficiency $\alpha_{\rm CE}$ = 1.0 and Maxwellian distribution natal
 kicks with $\sigma=265\kms$ \citep{Hobbs2005} lowered by fallback \citep{Fryer1204}. 
 We adopt solar metallicity Z=0.02 consistently 
with \cite{2017MNRAS.465.2092P}.

\begin{figure*}
   \centering
   \includegraphics[width=140mm]{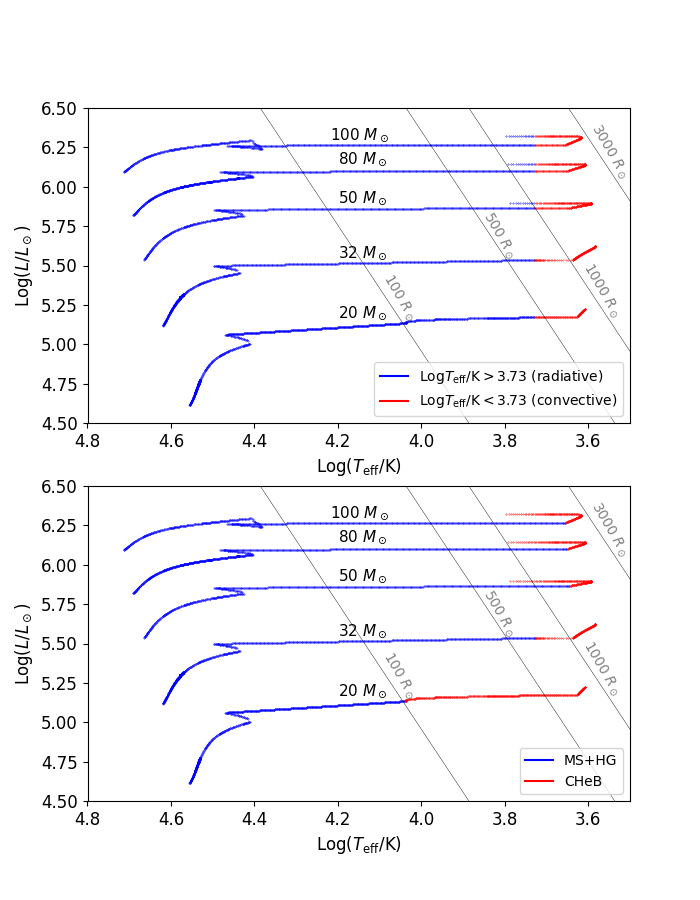}
   \caption{Hertzsprung–Russell diagram for massive stars with ZAMS masses: 
   $20\msun$, $32\msun$, $50\msun$, $80\msun$ and $100\msun$ at 
   metallicity Z=0.007; evolution of single star since main sequence to core 
   helium burning. In the top panel with the red line we marked stars cooler than 
   log($T_{\rm eff}$/K) $<$ 3.73  and with the blue line stars with log($T_{\rm eff}$/K) $>$ 3.73. It 
   is expected that the red line represents stars having convective envelopes while 
   the blue line is for radiative envelopes 
   \citep{2004ApJ...601.1058I,2008ApJS..174..223B}. In the bottom panel we marked 
   with the blue line star main sequence and Hertzsprung gap phases and with the 
   red line core helium burning phase. Evolutionary stages are defined as in 
   formulas by \cite{Hurley_2002} and \cite{2008ApJS..174..223B}. With the grey 
   lines we marked borders corresponding to stellar radii: 
   100$\rsun$, 500$\rsun$ and 1000$\rsun$. }
              \label{Fighg}%
    \end{figure*}

We assume that systems with Hertzsprung gap (HG) donor star merge during CE 
phase \citep{2007ApJ...662..504B}. In {\tt StarTrack} code HG phase begins 
after leaving the main sequence and for both less and more massive stars it 
is the period of intense star expansion (as during the
main sequence stellar radius usually does not increase more than by a factor of few). Therefore
during HG phase stars often initiate TTMT and CE. At the onset of RLOF such donors are 
often only partially expanded post main sequence stars.
Therefore, it is not well known whether such objects have already well-separated 
core and envelope structure.
In Figure \ref{Fighg} we present Hertzsprung–Russell evolution diagram for 
massive single stars. On the top panel of the figure we marked part of the 
evolution when stars are expected to have radiative or convective envelopes. 
On the bottom panel we marked part of evolution where stars are defined in 
{\tt StarTrack} as main sequence+HG or core helium burning.

We tested delayed supernovae (SN) engine \citep{Fryer1204,Belczynski1209} which
affect the birth mass of NSs and BHs, allowing for the formation of the 
compact objects within the first mass gap ($\sim 2-5\msun$). 
We assume that maximum mass of the NS is 
2.5$\msun$ \citep{MaxNS} so more massive compact objects are BHs. 

In this work we present results for three different evolutionary models: a model with standard
{\tt Startrack} treatment of RLOF: M380.B and the
other two which include the revised RLOF treatment: M480.B and M481.B. In the names
"M" stands for a model, then the number is increasing with time to mark subsequent physical 
models. The ending "B" refers to submodel B often used in previous {\tt Startrack} works 
e.g., \cite{2020A&A...636A.104B}. In submodel B all HG donor systems merge during the CE phase.
Models are listed in Table \ref{tab:model_tested}.

   \begin{table} 
      \caption[]{List of tested evolutionary models.}
        \label{tab:model_tested}
     $$ 
         \begin{array}{lccc}
            \hline
            \noalign{\smallskip}
            \mbox{Model} & \mbox{CE development criteria} & \mbox{MT switch}\\
            \noalign{\smallskip}
            \hline
            \hline
            \noalign{\smallskip}
            \mbox{M380.B} & \mbox{Standard} & \mbox{Standard} \\
            \mbox{M480.B} & \mbox{Revised} & \mbox{Standard}  \\
            \mbox{M481.B} & \mbox{Revised} & \mbox{Modified} \\
            \hline
         \end{array}
     $$ 
   \end{table}
\color{black}

\section{RLOF} \label{sec: ce_procedure}

Two new types of instabilities were identified in RLOF binaries with massive donors, 
if such donors were allowed to evolve while they are exceeding their Roche lobe 
-- the expansion and the convective instability \citep{2017MNRAS.465.2092P}.
Expansion instability happens if an RLOF donor experiences a  period of fast 
thermal-timescale expansion after its main sequence. This expansion may lead 
to the development of dynamical instability in a very short time 
(a few thousand years) after the start of thermal time-scale mass transfer (TTMT). 
The second type of instability, convection instability, is associated 
with developing a sufficiently deep convective envelope.

In this work, we implement in our population synthesis studies the revised 
development criteria for the occurrence of the two instabilities and examine 
how they influence the formation of DCO mergers.  We base our criteria for 
whether any of the instabilities occur in a given donor on the numerical results 
of  \cite{2017MNRAS.465.2092P}, who reported the boundaries between the 
instabilities using the radii of the donors of different masses and metallicities 
at the onset of RLOF.
Following \cite{2017MNRAS.465.2092P} we define two radii:  $R_{\rm U}$- smallest 
radius for which the convection instability occurs, and $R_{\rm S}$- the maximum 
radius when the expansion instability can take place. 

In our approximation, the values of $R_{\rm U}$ and $R_{\rm S}$ are the averages of 
the ranges obtained and given in Table 1 of \cite{2017MNRAS.465.2092P}. The specific 
values of $R_{\rm U}$ and $R_{\rm S}$ used in this work are given in Table \ref{tab:natasha}.  
Note that simulations by \cite{2017MNRAS.465.2092P} were performed for the systems 
with BH companion. In our work, we extend the results and treat the same way binary 
systems with any other companion types (e.g., main-sequence stars, neutron stars, etc.). 
If radius $R_{\rm don}$ of a given donor at the onset of RLOF is found within 
range $R_{\rm S}$-- $R_{\rm U}$ for $Z \leq 0.5 Z_{\odot}$ or below 
$R_{\rm don}$<$R_{\rm U}$ for $Z > 0.5 Z_{\odot}$ then we assume stable RLOF, 
otherwise CE evolution is applied. 
Note that stars of a higher metallicity were not found to experience the expansion 
instability \citep{2017MNRAS.465.2092P}.

   \begin{table} 
      \caption[]{Boundary radii values between the stable and unstable RLOF,
      values based on data 
      from \cite{2017MNRAS.465.2092P}. If radius $R_{\rm don}$ of a given donor during
      the ongoing RLOF is found within range 
      $R_{\rm S}$-- $R_{\rm U}$ for $Z \leq 0.5 Z_{\odot}$ or below 
      $R_{\rm don}$<$R_{\rm U}$ for $Z > 0.5 Z_{\odot}$ then we assume stable 
      RLOF, otherwise CE evolution is applied. There are cases where expansion 
      instability was not found for any radius (stable) and where cases where RLOF 
      is unstable over whole radius range (unstable).}
        \label{tab:natasha}
     $$ 
         \begin{array}{cccc}
            \hline
            \noalign{\smallskip}
            {M}_{\rm{don}} [{M}_{\odot}]     &  {M}_{\rm{comp}} [{M}_{\odot}] & {R}_{\rm{S}} [{R}_\odot] & {R}_{\rm{U}} [{R}_\odot]\\
            \noalign{\smallskip}
            \hline
            \hline
            \noalign{\smallskip}
            \multicolumn{3}{c}{\mbox{Metallicity $Z$ $\leq$ 0.5 $Z_{\odot}$}}    \\
            \hline
            19.6 & 7 & {\rm stable} & 703.5\\
            29.1 & 7 & 47.5 & 1057.5 \\
            37.6 & 7 & 331.5 & 1293.5 \\
            56.8 & 7 & \multicolumn{2}{c}{\rm unstable} \\
            56.8 & 10 & 355.0 & 1747.5 \\
            56.8 & 12 & 148.0 & 1823.5\\
            74.5 & 7 & \multicolumn{2}{c}{\rm unstable} \\
            74.5 & 10 & {\rm stable} & 2229.0\\
            74.5 & 14 & 144.5 & 2150.5\\
            \noalign{\smallskip}
            \hline
            \noalign{\smallskip}
            \multicolumn{3}{c}{\mbox{Metallicity $Z$ > 0.5 $Z_{\odot}$}}    \\
            \hline
            19.6 & 7 & {\rm stable} &736.0 \\
            26.6 & 7 & {\rm stable} & 1159.0 \\
            32.5 & 7 & {\rm stable} & 1407.5 \\
            41 & 10 & {\rm stable} & 2103.5 \\
            41 & 12 & {\rm stable} & 2033.0 \\
            \hline
         \end{array}
     $$ 
   \end{table}

\subsection{CE development criteria} \label{stability}

In the revised treatment, we require the following four conditions to be met 
simultaneously for CE development:

\textbf{(i) The condition based on the donor type.} Revised RLOF instability 
criteria are applied if, during the RLOF, the donor is an H-rich envelope giant 
of one of the following types: HG star, first giant branch, core helium burning, 
early asymptotic giant branch, or thermally pulsing asymptotic giant branch. 
However, for HG donors, we assume that CE always leads to a merger, and we 
halt this system binary evolution (see \S \ref{sec: method}). 

\textbf{(ii) The condition based on the donor mass.}
Revised RLOF instability criteria are applied 
only to the systems with initially massive donor stars with 
$M_{\rm ZAMS,don}>18 \msun$.
 
\textbf{(iii) {The condition based on the system mass ratio.}} The ratio 
of companion masses $M_{\rm comp}$ to donor masses $M_{\rm don}$ 
(see Table \ref{tab:natasha}) represents the border between regimes of always 
stable and possibly unstable RLOF.
Therefore, from Table \ref{tab:natasha} we can obtain the limit value of mass ratio 
$q_{\rm {CE}}={M_{\rm comp}}/{M_{\rm don}}$ for binary to enter potentially 
unstable RLOF regime.
For metallicities $Z \leq $ 0.5 $Z_\odot$ and for different donor mass $M_{\rm{don}}$ ranges,
${q_{\rm {CE}}}$ values are following:
\begin{equation}
\label{eq:1}
\begin{split}
     {q_{\rm CE}} = 0.36  \hspace{0.5cm}  \rm{ for } \hspace{0.5cm}  \textit{M}_{\rm{don}} \in (18,60) \textit{ M}_{\odot} \\
     {q_{\rm CE}} = 0.21  \hspace{0.5cm}  \rm{ for } \hspace{0.5cm}  \textit{M}_{\rm{don}} \in [60,80) \textit{ M}_{\odot} \\
     {q_{\rm CE}} = 0.19  \hspace{0.5cm}  \rm{ for } \hspace{1.2cm}  \textit{M}_{\rm{don}} \geq 80 \textit{ M}_{\odot}
\end{split}
\end{equation}
while for metallicities $Z > 0.5$ $Z_\odot$ 
\begin{equation}
\label{eq:2}
\begin{split}
     {q_{\rm CE}} = 0.36 \hspace{0.5cm}  \rm{ for } \hspace{0.5cm}  \textit{M}_{\rm{don}} \in (18,60) \textit{ M}_{\odot} \\
     {q_{\rm CE}} = 0.29 \hspace{0.5cm}  \rm{ for } \hspace{0.5cm}  \textit{M}_{\rm{don}} \in [60,80) \textit{ M}_{\odot} \\
     {q_{\rm CE}} = 0.19 \hspace{0.5cm}  \rm{ for } \hspace{1.2cm}  \textit{M}_{\rm{don}} \geq 80 \textit{ M}_{\odot}
\end{split}
\end{equation}

If the mass ratio of the binary system is greater than $q_{\rm CE}$, we
assume stable RLOF. 

\begin{figure*}
   \centering
   \includegraphics[width=140mm]{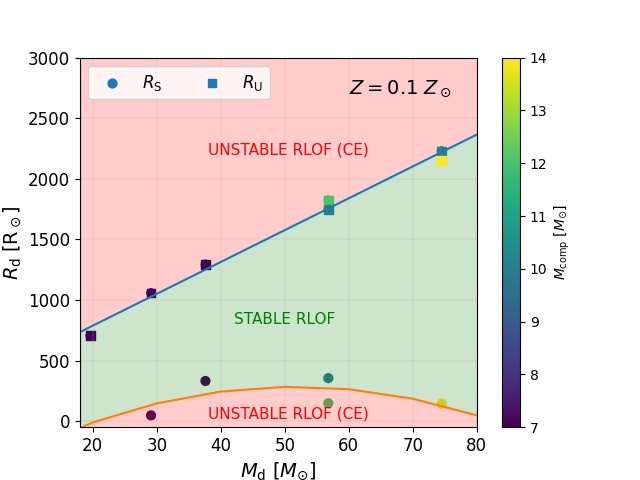}
   \caption{
Revised CE development criteria for stars with $10\%$ of solar metallicity. On 
horizontal axis is the mass of the donor star. On the vertical axis is the 
radius of the donor. The circle and square points mark values of 
$R_{\rm S}$ and $R_{\rm U}$, respectively (see Table~\ref{tab:natasha}). 
We fit lines to these data 
(see eq.~\ref{eq.low_metallicity1},\ref{eq.low_metallicity2} and 
\ref{eq.high_metallicity1}) to show regions of stable RLOF and unstable RLOF
(CE). The color coded bar on the right denotes companion mass.}
\label{FigDia2}%

   \includegraphics[width=140mm]{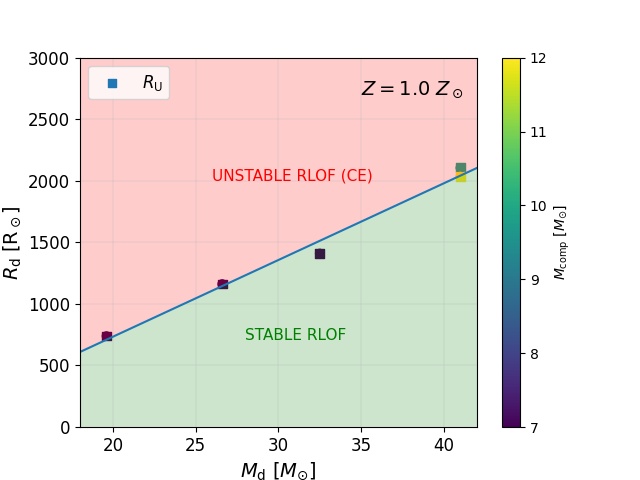} 
   \caption{
Revised CE development criteria for stars with $100\%$ of solar metallicity.
Notation is the same as on Figure~\ref{FigDia2}.}
              \label{FigDia1}%
    \end{figure*}

\textbf{(iv) The condition based on the donor radius.} 
Finally, if all conditions i-iii are met, we use stability diagrams presented in
Figures \ref{FigDia2} and \ref{FigDia1} to decide between stable RLOF and CE development. Diagrams 
are for two metallicities corresponding to simulations performed by \cite{2017MNRAS.465.2092P}: 
$Z=0.1$ $Z_\odot$ and $Z=1.0$ $Z_\odot$  respectively. We marked the values of $R_{\rm U}$ and $R_{\rm S}$
with the corresponding donor masses taken from the Table \ref{tab:natasha} and fit functions 
to the simulation points. We extend our model for wider metallicity ranges and follow different 
procedures for metallicities smaller than 0.5 $Z_\odot$ and greater than 0.5 $Z_\odot$. \\
For $Z \leq$ 0.5 $Z_\odot$ we used simulation data for 
$Z = 0.1$ $Z_\odot$ and we fit second-degree polynomial to $R_{\rm S}$ points with 
the following coefficients: 
\begin{equation} \label{eq.low_metallicity1}
    {R_{\rm{S}} = -0.29 M_{\rm{don}}^2+30.3 M_{\rm{don}}-498}
\end{equation}
and a straight line to $R_{\rm U}$:
 \begin{equation} \label{eq.low_metallicity2}
   {R_{\rm{U}} = 26.3 M_{\rm{don}}+262.}
\end{equation}
If a donor during ongoing RLOF has a radius 
$R_{\rm don}$ $<$ $R_{\rm S}$ or $R_{\rm don}$ $>$ $R_{\rm U}$, RLOF is 
unstable (CE). \\
For $Z>$ 0.5 $Z_\odot$ we used simulation data for 
$Z = 1.0$ $Z_\odot$. In this case, we had only $R_{\rm U}$ points and 
we fit a straight line with the equation: 
\begin{equation} \label{eq.high_metallicity1}
   {R_{\rm U} = 62.3 M_{\rm don}-515.} 
\end{equation}
If a donor radius during ongoing 
RLOF is $R_{\rm don}$ $>$ $R_{\rm U}$, RLOF is unstable (CE).\\ 
$M_{\rm don}$ in Equations \ref{eq.low_metallicity1}-\ref{eq.high_metallicity1} stands for 
the mass of the donor at a given RLOF time step, such that only 1\% of donor mass is 
transferred (see \S 5.1, \cite{2008ApJS..174..223B}).

In detailed simulations, RLOF is always unstable if $q < 0.123 $ (see also Table 
\ref{tab:natasha}).  We therefore adopt that CE event always starts in binaries 
where at the onset of RLOF the mass ratio is
\begin{equation} 
 \frac{M_{\rm comp}}{M_{\rm don}}<q_{\rm crit}=0.125\ . 
\end{equation}

Unfortunately, He giant donors do not have envelopes similar to hydrogen stars 
to make a straightforward connection and use revised RLOF stability diagrams. 
Therefore, in this work for He stars as well as in cases when the 
donor is a main sequence star (H-rich; core H-burning and He-rich; 
core He-burning main sequence) or a giant of $M_{\rm ZAMS,don}< 18 \msun$ 
we follow the standard {\tt StarTrack} CE 
development criteria and stable RLOF treatment \citep{2008ApJS..174..223B}.

\subsection{Stable RLOF} \label{timescales}

In this study, we also test a modification of the condition for switching from TTMT 
to nuclear-timescale stable MT in the {\tt StarTrack} code.
So far, once the TTMT began, the issued MT was continued until the donor star's 
radius $R_{\rm don}$ decreased below 1.1 of its Roche lobe radius $R_{\rm lobe}$.
Then the type of MT was selected based on the comparison of 
MT timescale  $\tau_{\rm eq}$ 
\citep[Equation 46 from][]{2008ApJS..174..223B} and thermal 
timescale $\tau_{\rm thermal}$. For $\tau_{\rm thermal}$ we use the approximation 
from  \cite{1996ApJ...458..301K}:

\begin{equation}
\tau_{\rm thermal} \approx 30 \left(\frac{M_{\rm don}}{\msun}\right)^2 \left(\frac{R_{\rm don}}{\rsun}\right)^{-1}\left(\frac{L_{\rm don}}{\lsun}\right)^{-1}{\rm Myr} \ .
\end{equation}

\noindent If $\tau_{\rm eq}>\tau_{\rm thermal}$, it was assumed that MT is driven 
on nuclear timescale. If $\tau_{\rm eq} \leq \tau_{\rm thermal}$, it was adopted 
the MT proceeds on thermal timescale.

In one of the tested models of this work, we modified this condition allowing
for the switch from TTMT to nuclear timescale stable MT always based on the 
timescale comparison in the given RLOF time step (not only if $R_{\rm{don}}<1.1 R_{\rm lobe}$). 
Note that this modification is applied to all binaries during their RLOF, 
regardless of components masses or types (as opposed to the new CE development 
criteria described in the \S \ref{stability}).

The evolutionary formulas we employ to evaluate star properties (e.g., radii,  mass, luminosity) 
based on \cite{2000MNRAS.315..543H} are not applicable for thermal-timescale changes of the 
star outside of thermal-timescale evolution that is intrinsic to normal single stars 
(e.g., post-main-sequence expansion). Therefore, we can not properly evaluate the donor's 
properties during ongoing TTMT, as then the donor is out of its thermal equilibrium.  
We obtain TTMT rate using mass, radius, and luminosity that the donor had when the TTMT 
just had started. Hence, the donor is still in thermal equilibrium. 
Therefore, we assume that the donor's properties should still be approximated well by our 
parameterized equations.
This obtained TTMT rate is kept constant for as long as TTMT proceeds. 
We have to resort to this simplified way to obtain MT rate as, while TTMT proceeds, our 
evolutionary formulas generate unreliable star radius and luminosity.

Yet, we use the onset RLOF parameters to construct our switches. In the first 
(original StarTrack) switch from TTMT to nuclear timescale MT, we wait until the donor's 
radius, as given by the evolutionary formula, becomes comparable to the size of the 
donor's Roche lobe. Then we assume that the star is back (or close) to its thermal equilibrium. 
Then we use the new donor's properties to decide whether to continue TTMT (with a new MT 
rate based on the  current donor mass, radius, and luminosity), or change to nuclear 
timescale MT. 

In the modified approach that we test in this work, at each time step of the ongoing TTMT 
we use the evolutionary formulae and calculate the current thermal and nuclear timescales. 
Suppose we get that the updated nuclear timescale is longer than the updated thermal 
timescale and the donor's radius is still larger than its Roche lobe radius. In that case, 
we assume that the donor has regained its thermal equilibrium, and we proceed with the MT 
on the new nuclear timescale. 
This usually happens earlier in the ongoing TTMT than meeting the original condition 
$R_{\rm don}<1.1R_{\rm lobe}$. 
With this new additional condition, a numerical artifact is plausible: the donor star's 
radius may significantly exceed its Roche lobe when we change MT to nuclear timescale. 
In this case we apply a ``safety" condition such that if 
\begin{equation}
R_{\rm don} > 2.0 R_{\rm lobe}
\end{equation}
\noindent we proceed with CE evolution for giant donors, or assume merger for main 
sequence donors (both  H-core burning H-rich and He-core burning He-rich main sequence stars). 

Using either the original switch or the modified one is not perfect. Since we cannot be 
sure which transition condition from TTMT to nuclear timescale MT would correspond better 
to detailed MT calculations, we will present models with both treatments. 

\subsection{Other cases of unstable RLOF} \label{other_ce}

There are other situations when RLOF is unstable, leading to a CE phase. We take 
into account the following two of the known scenarios: {\em (i) } 
in case of the Darwin instability, due to an extreme mass ratio \citep{1993ApJS...88..205L};
{\em (ii)} if the accretion flow is so strong that photons are trapped in it, building up an 
envelope around a compact accretor in excess of its Roche lobe radius  \citep{1979MNRAS.187..237B}. 
For these, we use standard {\tt StarTrack} procedures \citep{2008ApJS..174..223B}. \\

\section{Results}

\subsection{NS-NS, BH-NS and BH-BH local merger rate density} \label{sec:results_rates}
We present local merger rate densities ($z \sim 0$) for different types 
of DCO systems corresponding to the tested physical models with standard (M380.B) 
and revised RLOF treatment (M480.B, M481.B) listed in the Table \ref{tab:rates}. 
Our rates are placed together with values recently estimated by LIGO/Virgo 
(90\% credible limits): $23.9^{+14.9}_{-8.6} \gpy$ for BH-BH mergers, 
${320^{+490}_{-240}}$ $\gpy$ for NS-NS mergers  \citep{catalog03} and previously 
given range for BH-NS mergers: $0-610 \gpy$ \citep{Abbott_2019}.

The effect of revised CE development criteria (separate from other tested changes) 
on the merger rates is visible by comparing results of model M380.B and M480.B. 
It seems to be rather counterintuitive as the rate for BH-BH and BH-NS mergers 
increased under more restricted conditions for CE initiation. In standard model (M380.B)
the local merger rate densities were for BH-BH: 62 $\gpy$ and for BH-NS: 13 $\gpy$  
while in M480.B they slightly increased to 88 $\gpy$ and 16 $\gpy$ respectively. 
This is caused by the fact that in the revised treatment during the RLOF some of 
the systems with HG star donor instead of entering the unstable RLOF regime (which 
always lead to the merger, \S {\ref{sec: method}}), go through the period of TTMT
which allows for further binary evolution and potential formation of 
BH-BH systems. On the other hand, systems initiate CE phase much less often and as 
the consequence we obtain new, dominant evolutionary scenario leading to formation 
of BH-BH mergers without any CE phase. In other words, the early (HG donors) development
of CE during TTMT eliminates many progenitors of BH-BH systems from dominant 
formation channel of model M480.B and the restricted (in terms of mass and mass ratio) 
development of CE reduces number of BH-BH mergers from the dominant formation 
channel in model M380.B. For details about the new evolutionary 
scenario see \S 5.  
The rate for NS-NS mergers (model M480.B), $\sim 150 \gpy$, is the same as in 
the standard model (M380.B), as the revised CE development criteria were adopted 
only to the systems with initially massive donors ($M_{\rm{ZAMS}}>18 \msun$, 
see \S \ref{stability}) which are mainly BH progenitors. Therefore NS-NS 
systems formation was not significantly influenced and the merger rate is 
within LIGO/Virgo range as it used to be for the standard approach. 
Rates for BH-NS systems for both M380.B and M480.B models are also consistent with the given LIGO/Virgo
range, which is wide due to the limited detection data for BH-NS mergers \citep{catalog03}. 
BH-BH merger rates for the same two models are about 1.5-2.0 times larger than 
the upper limit given by LIGO/Virgo.

Model M481.B includes both implemented changes to our standard approach (M380.B): 
revised CE development criteria and modified condition for switch between TTMT 
and nuclear timescale stable MT. As the second change was implemented to all systems 
(regardless components masses or types) it affected also NS-NS merger rate 
(for M481.B equal $\sim 320 \gpy$) increasing it about 2 times comparing with the 
M380.B and M480.B models. 
The additional modification implemented in model M481.B leads, however, to decrease 
of the rate for BH-BH and BH-NS mergers (18 $\gpy$ and 4 $\gpy$ respectively)
by a factor of $\sim 4-5$ comparing to the model M480.B. 
The effect described in previous paragraph which leads to the increase of the survived BH 
binaries progenitors number due to entering stable RLOF instead of early CE with HG donors,
does not apply here. 
Due to the modification of condition for switch between MT types, TTMT stops
faster than in the standard approach. At the same time donor's radius derived from the formula 
for stars in equilibrium (see \S \ref{timescales}) is large ($R_{\rm don}>2.0 R_{\rm lobe}$) 
what leads to early CE initiation (and merger of all HG donor systems). The new dominant 
BH-BH merger formation channel emerges in this model and it is different than dominant BH-BH 
formation channels in models M380.B and M480.B.
This channel includes TTMT that develops into late CE with core helium burning donor (for details 
see \S 5). Merger rates for all types of DCOs in this (M481.B) model are in good agreement 
with recent LIGO/Virgo estimates.

   \begin{table} 
      \caption[]{Local ($z \sim 0$) merger rate densities 
      $\left[\mbox{Gpc}^{- 3} \mbox{yr}^{- 1} \right]$ for different types of 
      DCO systems. With the bold font we mark the model for which  BH-BH, BH-NS and 
      NS-NS merger rates are within ranges given by LIGO/Virgo 
      \citep{Abbott_2019,catalog03}.}
        \label{tab:rates}
     $$ 
         \begin{array}{lccc}
            \hline
            \noalign{\smallskip}
            \mbox{Model} & \cal{R}_{\mbox{\small{BH-BH}}}& \cal{R}_{\mbox{\small{BH-NS}}} & \cal{R}_{\mbox{\small{NS-NS}}} \\
             
            \noalign{\smallskip}
            \hline
            \hline
            \noalign{\smallskip}
            \mbox{LIGO/Virgo} & {23.9} & {-} & {320}\\
            \mbox{} & {15.3-38.8} & {0-610} & {80-810}\\
            \noalign{\smallskip}
            \hline
            \noalign{\smallskip}
            {\mbox{\bf{M481.B}}}  &\bf{17.9} &\bf{4.1} & \bf{322} \\
            {\mbox{{M380.B}}}  & {61.7} &{ 13.1} & {148} \\
            {\mbox{{M480.B}}} & {88.4} & {15.6} & {148}\\

            \hline
         \end{array}
     $$ 
   \end{table}
   
\subsection{BH-NS and BH-BH mass ratio distribution} \label{sec:results_mass_ratio}

Until the realise of O3 data, all ten detected O1/O2 BH-BH mergers published by LIGO/Virgo team 
were consistent  with  being  equal-mass merges \citep{LIGO2019a,
LIGO2019b,2020ApJ...891L..27F}. Latest detections indicated however, that the distribution of 
BH-BH mergers mass ratio may be more complex and broad. Recently two detections of DCO mergers
with highly asymmetric masses were announced \citep{gw190412,2020LIGOMassGap}. The events 
GW190412 ($\sim 30 \msun$ BH and $\sim 8 \msun$ BH) and GW190814 
($\sim 23 \msun$ BH and $\sim 2.6 \msun$ NS or BH) were caused by DCO mergers in which one of the objects 
was less massive than the second by a factor of $0.28_{-0.07}^{+0.12}$ and $0.112_{-0.009}^{+0.008}$ respectively.
We check how revised CE development criteria and stable RLOF treatment influence the 
mass ratio distribution of BH-BH, and BH-NS mergers, and if we are still able to 
reconstruct asymmetric mass BH-BH and BH-NS mergers as in previous models
\citep{olejak2020origin,Drozda2020}.
In our study, we define mass ratio of merging components $q$ as the ratio of less massive 
compact object to the more massive compact object in agreement with LIGO/Virgo definition.  

\begin{figure*}
   \centering
   \includegraphics[width=130mm]{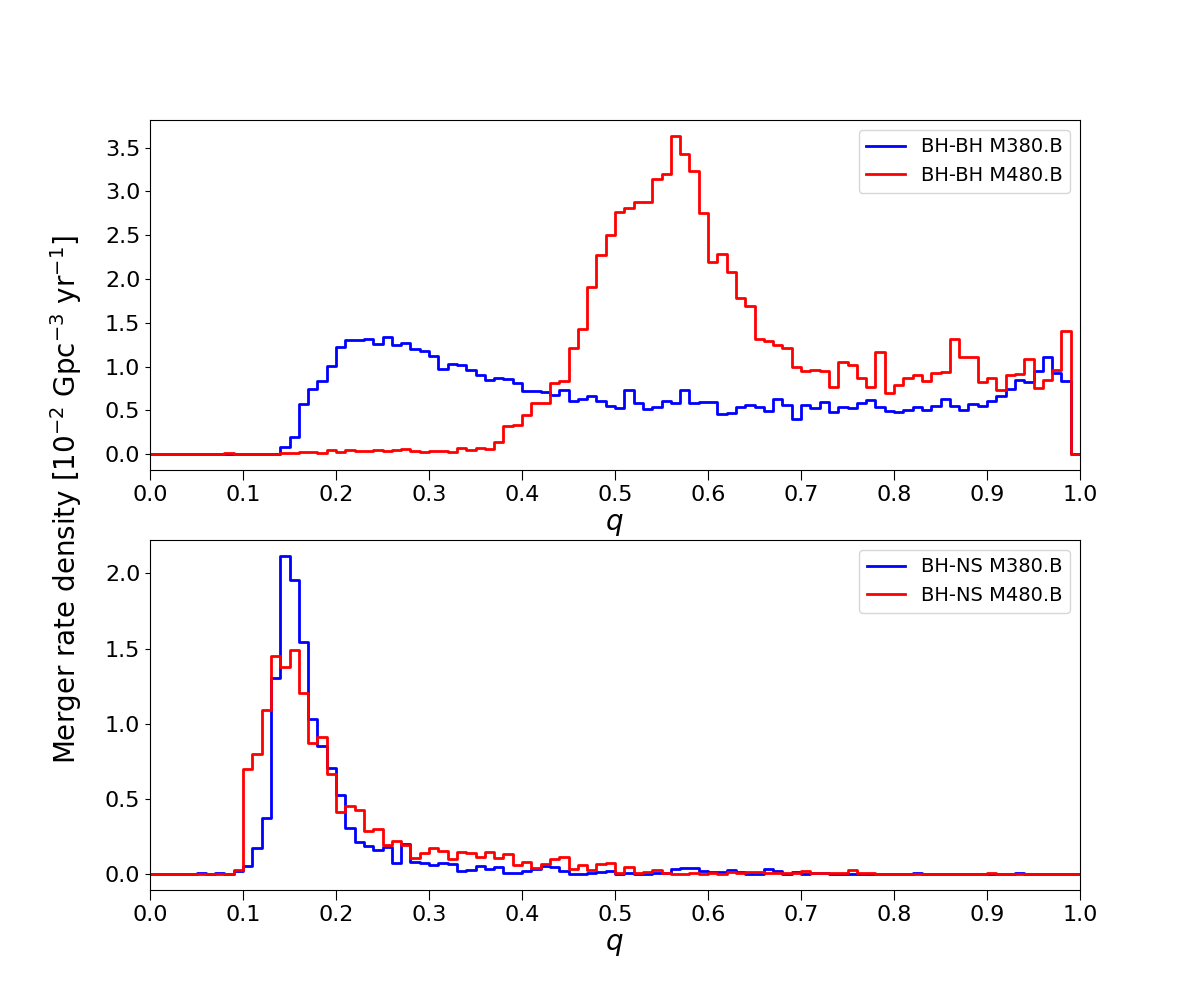}
   \caption{Distribution of mass ratio in BH-BH and BH-NS mergers 
   (z $\sim 0$). On the top panel 
   BH-BH mergers, on the bottom panel BH-NS mergers. Blue line - results for 
   standard CE development criteria with delayed SN engine (M380.B); red line - results 
   for revised CE development criteria (M480.B).}
              \label{Fig_q_ratio_thermal}%
   \centering
   \includegraphics[width=130mm]{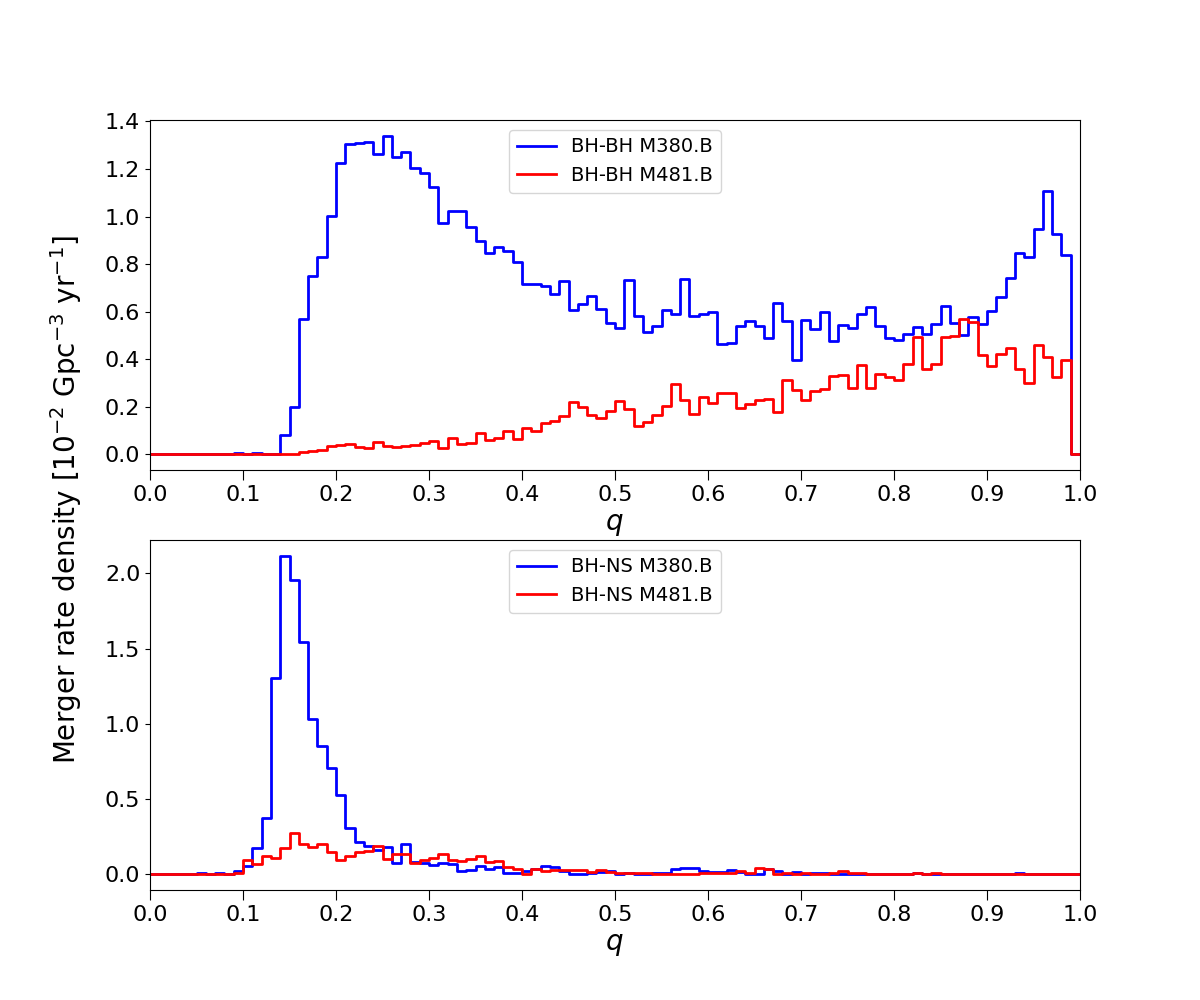}
   \caption{Distribution of mass ratio in BH-BH and BH-NS mergers 
   (z $\sim 0$). On the top panel BH-BH 
   mergers, on the bottom panel BH-NS mergers. Blue line - results for standard CE 
   development criteria with delayed SN engine (M380.B); red line - results for revised 
   CE development criteria and revised stable RLOF treatment (M481.B).}
              \label{Fig_q_ratio_nuclear}%

    \end{figure*}

In Figures \ref{Fig_q_ratio_thermal} and \ref{Fig_q_ratio_nuclear} we present 
distributions of mass ratio of BH-BH and BH-NS mergers ($z\sim 0$) for two models with revised 
RLOF treatment (red line): M480.B and M481.B respectively.

We calculated percent of asymmetric mass BH-BH mergers with $q<0.4$ which has 
been constrained after GW190412 detection to constitute 
$\gtrsim 10\%$ of overall BH-BH merger population \citep{gw190412}. In the standard 
$\tt StarTrack$ models with rapid (M230.B, \cite{olejak2020origin}) and delayed (M380.B) 
SN engine (\cite{Fryer1204}) those fractions are $9\%$ and $\sim 30\%$ respectively. 
We find that our tested revised CE development criteria reduced the fraction of unequal 
BH-BH mergers comparing to model M380.B. 
Percent of BH-BH mergers with $q<0.4$ for model M480.B constitute 
$\sim$ 2\% while for M481.B it is $\sim 6\%$. The main formation channel for the 
formation of unequal mass BH-BH binaries in standard approach 
(see Tab.1 of \cite{olejak2020origin}) is missing for the revised CE treatment 
(see Tab.4, \S 5). Systems that would produce unequal mass BH-BH mergers, 
in the revised treatment instead of CE, enter stable RLOF regime and finally form 
wide BH-BH binaries which do not merge in Hubble time. The lowest BH-BH merger 
mass ratio achieved in tested models is $q \sim 0.06$ for the M480.B and 
$q \sim 0.1$ for M481.B. 

In distribution of BH-BH mergers for model M480.B there is a
large peak in the range of $q \approx 0.4-0.6$ which is not present in 
distributions for other models. Vast majority (94\%) of BH-BH mergers in M480.B forms 
via evolutionary scenario without any CE phase (see Tab. \ref{tab:formation_channels} 
and Fig. \ref{BHBH_diagram}). Binary separation, instead during CE phase as in the standard 
{\tt Startrack} approach, is reduced as the donor mass is ejected from the system together 
with the orbital angular momentum during stable RLOF. Described mechanism of orbital 
shrinkage is effective only for unequal mass binary components, when donor mass at the 
RLOF onset is significantly larger than the companion's mass \citep{vandelHeuvel2017,Marchant2021}. 
In our simulations such systems usually end their evolution as BH-BH binaries with mass 
ratio close to $q=0.5$ (Fig. \ref{BHBH_diagram}).

Mass ratio distribution for BH-NS mergers is strongly dominated with low mass 
ratio systems, what has been already reported and described for example by 
\cite{Drozda2020}. For both models M480.B and M481.B, $\sim$ 90\% of the BH-NS mergers
have $q < 0.35$. However, the lowest achieved mass ratios for BH-NS merger ($\sim 0.1$) 
is not that extreme as it used to be in standard model M380.B (0.03).
Similarly as in case of the BH-BH system, the main formation channel leading to formation 
of extreme mass ratios (see Fig.6 in \cite{Drozda2020}) is reduced as during the first RLOF 
instead of CE systems goes through period of stable RLOF.

\subsection{BH-BH mass distributions } \label{sec:primary}

\cite{catalog03} gives recent estimations on the possible shape of mass distribution 
for the more massive of the merging BHs ($m_1$) based on so far detected BH-BH 
mergers. The proposed fit is a power-law mass function $p(m_1) \propto m_1^{-\alpha}$
with a break around $40 \msun$ with the first exponent for the masses 
$m_1 \lesssim 40 \msun$, $\alpha_1 = 1.58^{+0.82}_{-0.86}$ and the second exponent 
for $m_1 \gtrsim 40 \msun$, $\alpha_2 = 5.6^{+4.1}_{-2.5}$. 

\begin{figure*} 
   \centering
   \includegraphics[width=180mm]{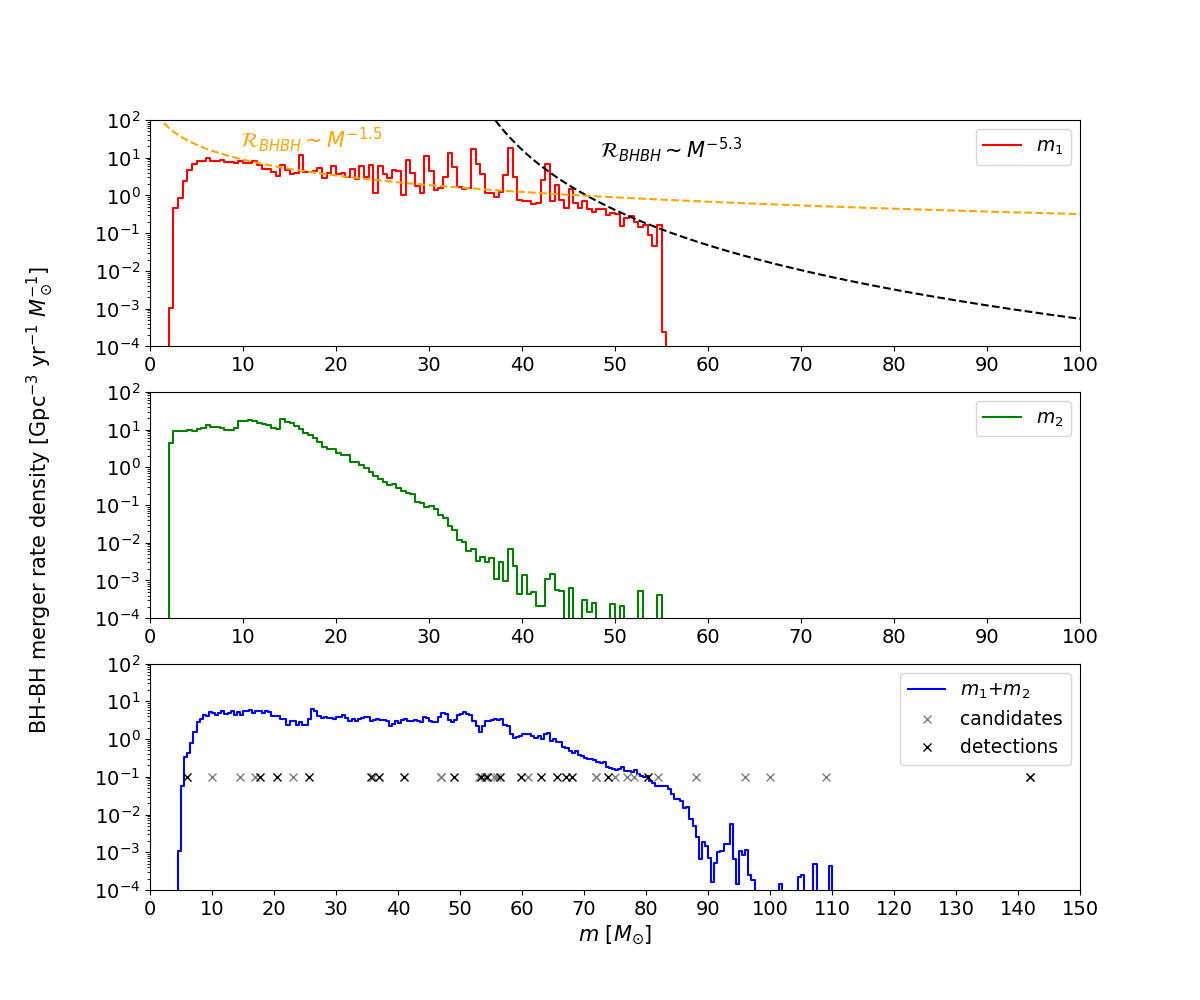}
   \caption{Intrinsic BH-BH merger ($z<2$) mass distributions: more massive BH mass (top
panel), less massive BH mass (middle panel), and total mass
(bottom panel). We also show total mass estimates for LIGO/Virgo BH-BH
mergers from O1, O2 and O3. Results for model M480.B. }
   \label{mass1_distribution_thermal}
    \end{figure*}

\begin{figure*} 
   \centering
   \includegraphics[width=180mm]{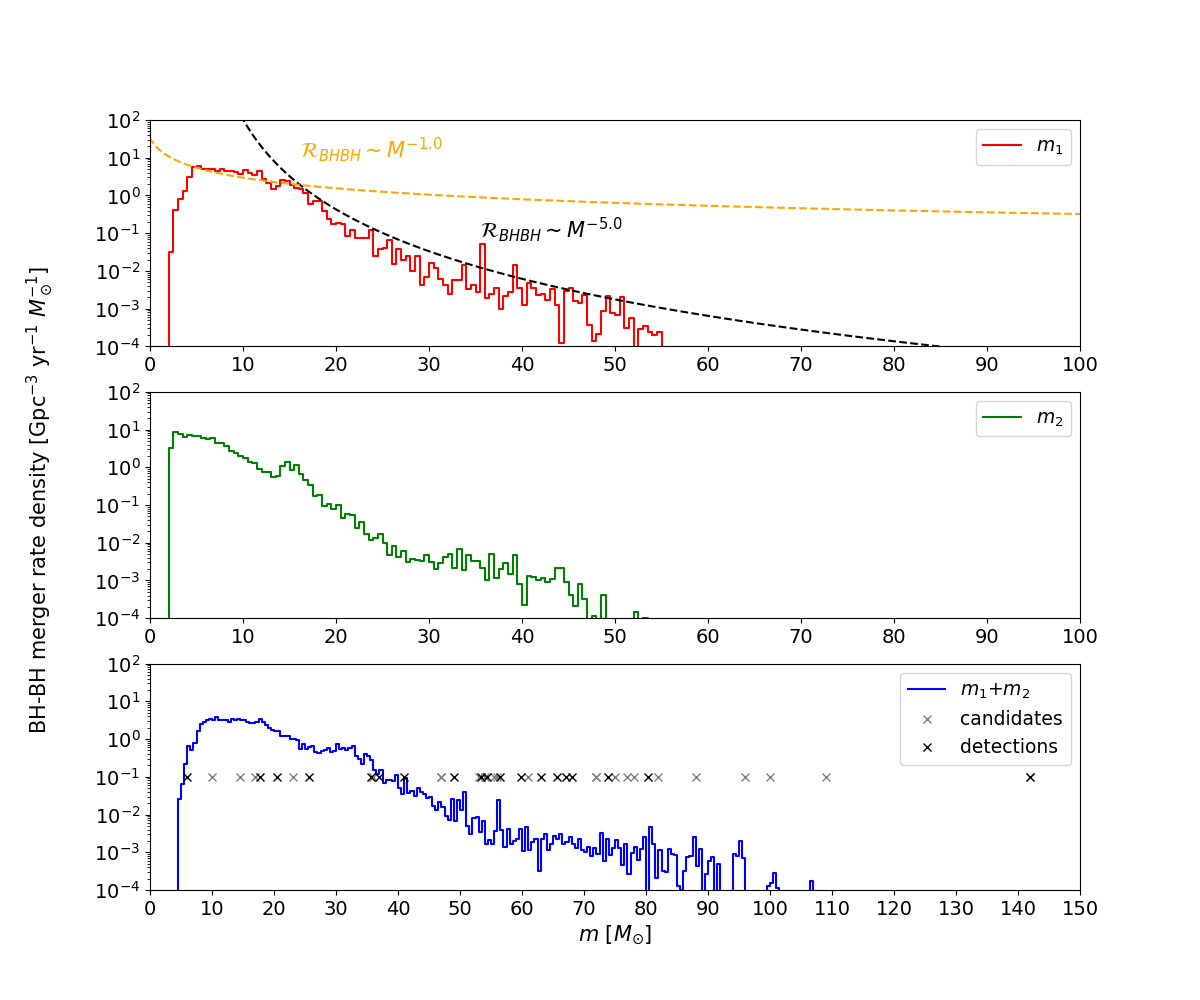}
   \caption{Intrinsic BH-BH merger ($z<2$) mass distributions: more massive BH mass (top
panel), less massive BH mass (middle panel), and total mass
(bottom panel). We also show total mass estimates for LIGO/Virgo BH-BH
mergers from O1, O2 and O3. Results for model M481.B. }
   \label{mass1_distribution_nuclear}
    \end{figure*}

We present the distributions of more massive, less massive BH masses 
and total mass of BH-BH systems which merged at redshifts $z<2$ for two tested models 
with revised RLOF physics. The Figures {\ref{mass1_distribution_thermal}} and 
\ref{mass1_distribution_nuclear} shows results for model M480.B and M481.B 
respectively. We approximate more BH mass distributions (Figs. \ref{mass1_distribution_thermal} 
and \ref{mass1_distribution_nuclear}; top
panels) with power-law function for an easy visual comparison with the LIGO/Virgo 
estimate.

The shape of more massive BH distribution for our model M480.B is in a very good 
agreement with the shape of the BH-BH detections (Fig. \ref{mass1_distribution_thermal}, 
top panel). The distribution for this model may be divided as well into two parts, 
with the break around $45 \msun$. 
The first part, up to the break is characterized by a slow decrease corresponding 
to the exponent $\alpha_1 \approx 1.5$ while after the break, the curve become much 
steeper with $\alpha_2 \approx 5.3$ and the cut at $\sim 55 \msun$ 
(PSN, begining of the second mass gap). 
Note that this model distribution is a better match to the LIGO/Virgo estimate  
than recently published {\tt StarTrack} results (see Fig.15 for model M30.B in
\cite{2020PhRvL.125j1102A}: single power-law with index $\alpha=3.6$. The LIGO/Virgo total 
BH-BH system masses are within range of our simulated population of BH-BH mergers 
for model M480.B. The only exception is GW190521: the most massive detected BH-BH merger 
\citep{2020PhRvL.125j1102A}, that has both BHs (85$\msun$+66$\msun$) within our adopted
pair-instability  mass gap: $55-135\msun$ \citep{2020A&A...636A.104B}. 
However, even such massive BHs can be possibly produced by massive stars if 
uncertainties on nuclear reaction rates and mixing in stellar interiors are
taken into account \citep{2020ApJ...905L..15B}. 

In the case of model M481.B, the more massive BH mass distribution is not a good
match to LIGO/Virgo data. Although two power-law exponents ($\alpha_1 \approx 1.0$ and
$\alpha_2 \approx 5.0$) are within LIGO/Virgo estimates, the break appears at quite a low
mass $M_{\rm break} \approx 15-20 \msun$. Note also the total BH-BH mass distribution
ends at lower mass ($M_{\rm cutoff}=100 \msun$) as compared with model M480.B 
($M_{\rm cutoff}=110 \msun$).

\begin{figure*} 
   \centering
   \includegraphics[width=180mm]{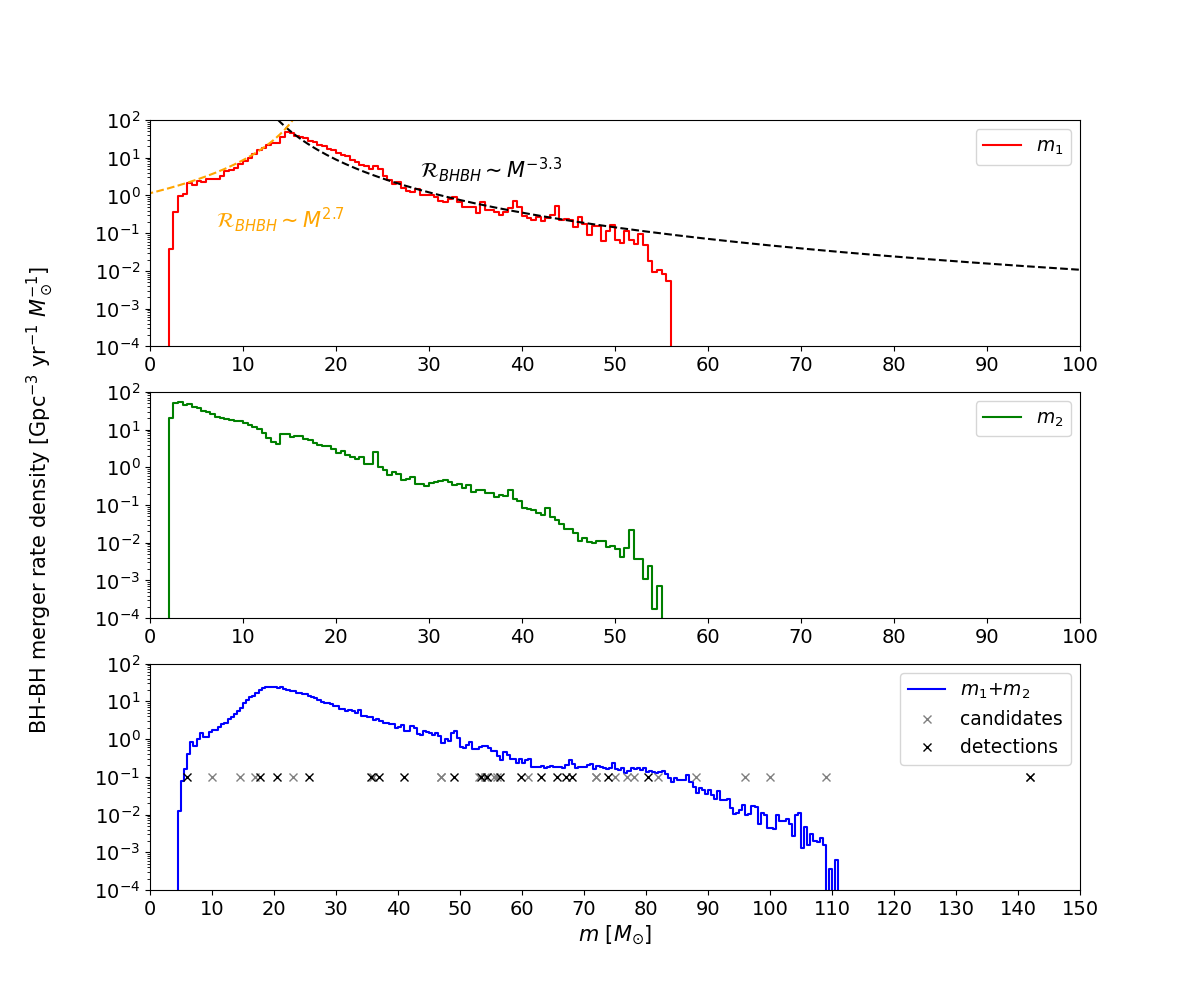}
   \caption{Intrinsic BH-BH merger ($z<2$) mass distributions: more massive BH mass (top
panel), less massive BH mass (middle panel), and total mass
(bottom panel). We also show total mass estimates for LIGO/Virgo BH-BH
mergers from O1, O2 and O3. Results for model M380.B (standard). }
   \label{mass1_distribution_standard}
    \end{figure*}

We also plot mass distributions for the model M380.B which represents our standard RLOF physics 
(Fig. \ref{mass1_distribution_standard}). Similarly to M481.B, the break point of more massive 
BH mass distribution is at rather low BH masses, $M_{\rm break} \approx 15-20 \msun$. 
However, in this case for masses $m_1<M_{\rm break}$ the distribution is increasing
with BH mass ($\propto M^{+2.7}$), which is inconsistent with the LIGO/Virgo estimates. For masses 
$m_1>M_{\rm break}$ we fit power-law exponent $\alpha_2 = 3.3$. 
The maximum total mass ($m_1+m_2$) of the BH-BH merger for model M380.B is $110 \msun$.

Our results indicate that modifications in RLOF physics approach may drastically
influence the BH mass distribution of BH-BH mergers. Some approaches (M480.B) allow to match 
nicely the empirical data for more massive BH mass distribution while for others 
distribution shape is rather off (M380.B and M481.B).

\section{Formation scenarios} 

\subsection{BH-BH formation scenarios} \label{AppendixA}
\begin{figure*} 
   \centering
   \includegraphics[width=180mm]{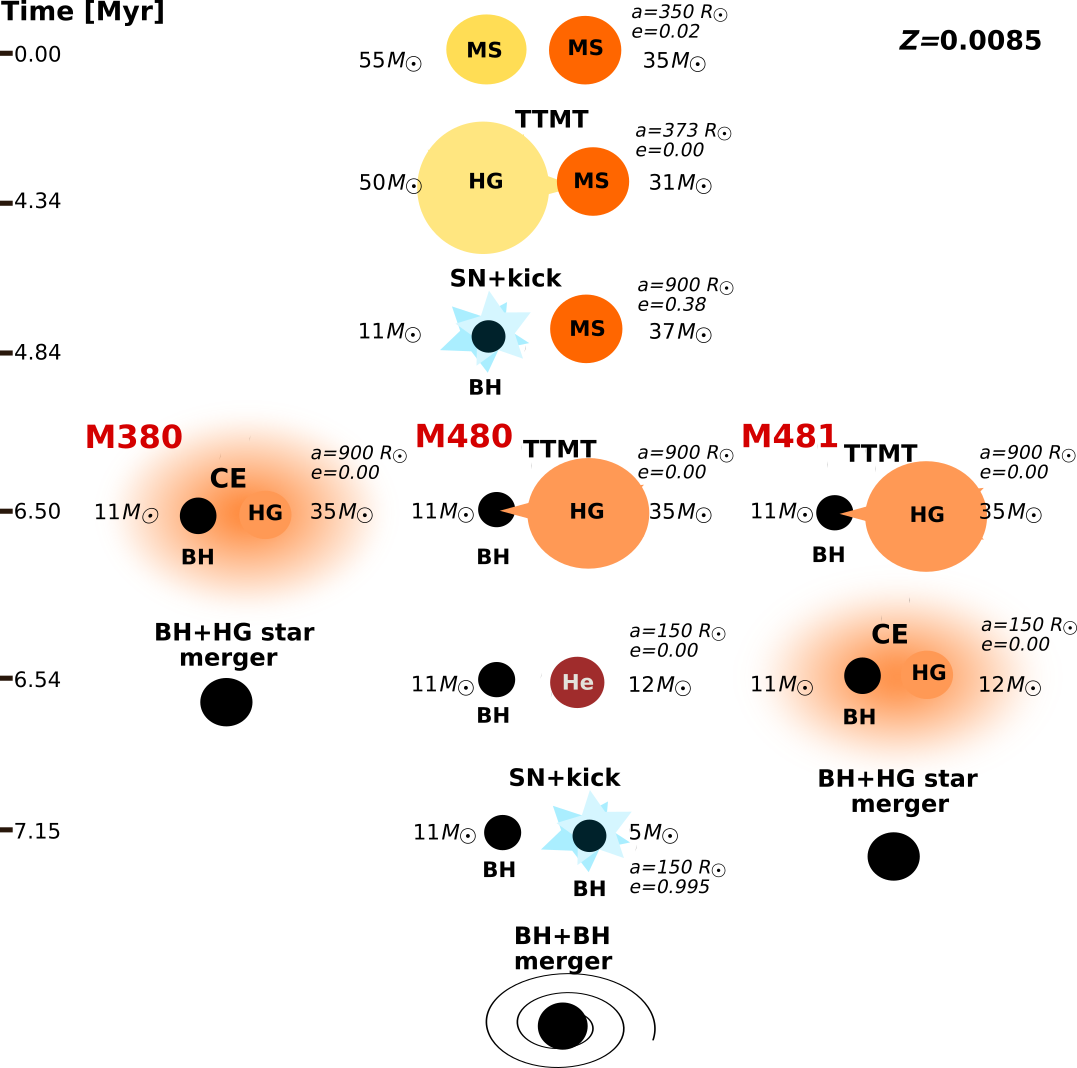}
   \caption{The typical local ($z \approx 0$) BH-BH mergers formation scenarios for 
   model M480.B together with the alternative evolution of the same system in models M380.B 
   and M481.B. For all models we begin with the same initial conditions, massive system of 
   55$\msun$ primary and 35$\msun$ secondary on the orbit of 350$\rsun$ and with metallicity 
   $Z=0.0085$. In all scenarios the primary initiates TTMT when it leaves main sequence and 
   begins to expand in thermal timescale during the HG phase. After 4.8 Myr the primary 
   finishes evolution and explodes as type Ib/Ic SN leaving behind a 11$\msun$ BH remnant. 
   Next, after the secondary leaves main sequence, due to more restricted condition for CE 
   development, in the cases of model M480.B and 481.B system goes through TTMT instead of 
   CE phase as in standard M380.B model. During TTMT the secondary looses large percent of 
   its mass (over 60\%) together with the system orbital angular momentum. Therefore, the 
   orbit tightens (by a factor of $\sim 4$). After TTMT, in model 480.B the system remains 
   BH-He system on the orbit of $150 \rsun$. Due to the high natal kick after the second SN 
   explosion, the orbital eccentricity significantly increases to e=0.995 what allows the BH-BH 
   system to merge in Hubble time.
   In standard (M380.B) scenario systems goes through CE phase with HG donor and we assume system
   a merger (see \S \ref{sec: method}).
   In model M481.B due to modified condition for switch between the TTMT and nuclear-timescale 
   stable MT, and the "safety" condition (read \S \ref{timescales}) after TTMT system also enters 
   CE phase with HG donor and merges leaving behind a single BH. "He" in the diagram stands for 
   a stripped helium core.}
   \label{BHBH_diagram}
    \end{figure*}

\begin{figure*} 
   \centering
   \includegraphics[width=180mm]{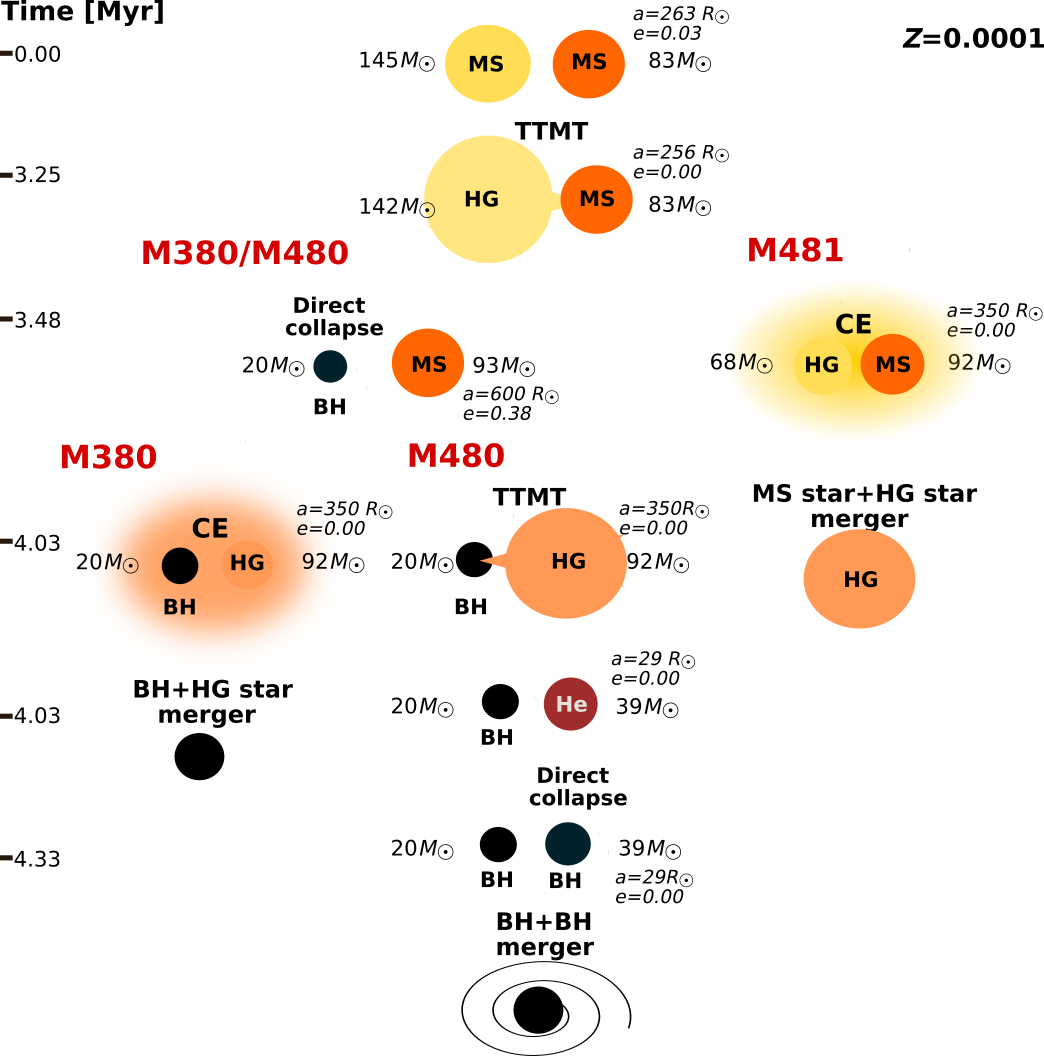}
   \caption{The typical local ($z \approx 0$) BH-BH mergers formation scenarios for 
   model M480.B together with the alternative evolution of the same system in models M380.B 
   and M481.B. For all models we begin with the same initial conditions, massive system 
   of 145$\msun$ primary and 83$\msun$ secondary on the orbit of 263$\rsun$ and with 
   metallicity $Z=0.0001$. In all scenarios the primary initiates TTMT when it leaves 
   main sequence and begins to expand in thermal timescale during the HG phase. When 
   TTMT finishes, in case of models M380.B and M480.B secondary looses envelope. A 
   BH with stripped helium core companion on the orbit of $350 \rsun$ is left. In 
   model M481.B due to modified condition for switch between the TTMT and nuclear-timescale 
   stable MT, and the "safety" condition (read \S 3.2) when TTMT ends system goes through CE 
   phase. As the donor is HG star, the CE ends with the system merger leaving behind 
   unevolved single star.
   After 3.3 Myr in models M380.B and M340.B the primary finishes evolution and collapse 
   to a 20$\msun$ BH remnant. Next, after the secondary leaves main sequence, due to more 
   restricted condition for CE development, in cases of model M480.B system goes through 
   TTMT instead of CE phase as in standard M380.B model. At the onset of TTMT the secondary 
   (donor) is over 4 times more massive than a BH companion. During the TTMT donor star 
   looses $\sim 60$ \% of its mass together with the system orbital angular momentum. This 
   leads to the significant orbit tightening by a factor of $\sim 12$. Efficient orbital 
   angular momentum loss during TTMT allows later formed 20 $\msun$ BH and $30 \msun$ BH 
   system to merge in Hubble time. In the scenario for M380.B highly unequal mass system enters 
   CE phase with HG donor. The binary system does not survive CE phase leaving behind a single BH.  
  "He" in the diagram stands for stripped helium core.}
   \label{BHBH_diagram_2}
    \end{figure*}

In Table {\ref{tab:formation_channels}} we provide evolutionary scenarios which lead to
the formation of local ($z\sim0$) BH-BH mergers for tested models with revised
CE development criteria (M480.B and M481.B) and standard $\tt StarTrack$ criteria 
(M380.B). For each model the dominant formation scenario, which constitutes the 
highest merger rate density fraction (marked with the bold text) is different. 
In Figures \ref{BHBH_diagram} and \ref{BHBH_diagram_2} we present diagrams
with the example for the dominant BH-BH formation scenario in model M480.B (without CE phase)
together with alternative evolution of the same system (the same initial conditions) 
but in models M380.B and M481.B. The Figure \ref{BHBH_diagram} is an example of a low-mass 
BH-BH merger formation, for which significant natal kick (received after second SN explosion) plays an important role. The second example (Fig. \ref{BHBH_diagram_2}) is massive BH-BH 
merger formation in which two BHs form via direct collapse.

For model M480.B (revised CE development criteria and standard switch from TTMT to 
nuclear-timescale stable MT) the dominant BH-BH mergers formation scenario is 
without any CE phase. Similar channels were reported in many other recent 
works \citep[]{vandelHeuvel2017,Andrews:2020pjg,zevin2020channel,Bavera2020,Marchant2021} 
as a possible formation scenario of LIGO/Virgo sources via isolated binary evolution.
The typical evolution of binary system in model M480.B is following: when primary 
(initially more massive star) leaves its main sequence and expands, system goes 
through a period of TTMT. Note that in some cases the donor star enters 
TTMT while is at its HG. Survival of a binary through this phase is facilitated by high TTMT rate from the donor star. This removes most of (or entire) donor's 
H-rich envelope and disallow significant increase of donor radius which in
turn does not allow for development of CE phase and possible merger of donor 
with its companion. After RLOF primary becomes a naked helium star. When primary star 
finishes evolution, the first BH is formed. In the meantime secondary evolves off main
sequence, expands and initiates a second phase of TTMT.  The second RLOF, as well as the first one,
is often initiated by a HG stars expanding on the thermal scales. If the system components 
have highly unequal masses (the donor mass is by a factor of 3 or more greater than the BH mass) 
the large mass loss from the system carries away also lots of the orbital angular momentum 
(even 90\% comparing with the amount at the second RLOF onset). This leads to significant orbit tightening. Such a mechanism has been reported 
previously e.g., by \cite{vandelHeuvel2017,Marchant2021}.
After MT, the secondary becomes naked helium star, and next forms a second, typically 
less massive BH. 

In all tested models BHs with masses $M_{\rm BH}$<10-15$\msun$ 
may receive a natal kick at the time of their formation (see \S \ref{sec: method}). The lower 
the BH mass, the greater probability of a high natal kick.
Therefore, in the BH-BH formation scenario the second-born BH (usually the less massive one) may 
receive a significant kick. High natal kick may impart significant 
eccentricity to newly formed BH-BH system what decreases the inspiral timescale
$T_{\rm ins} \sim (1-e^2)^{7/2}$,\citep[][]{1964PhDT........51P}, allowing such BH-BH system to merge in Hubble time. For example BH-BH system of 25$\msun$ and 10$\msun$ on the wide orbit of
$\sim$900$\rsun$ may merge in $\sim 10$ Gyr if the eccentricity is as high as $e=0.992$.
For a comparison, in the case of circular orbit ($e=0$) initial separation of the same BH-BH system 
would have to be less than $\sim 30\rsun$, for eccentricity $e=0.7$ less than $\sim$45$\rsun$ 
and for $e=0.9$ less than $\sim$110$\rsun$.
First stable RLOF phase in this scenario (M480.B) would be replaced in our standard model 
(M380.B) with CE phase for many binaries as revised CE 
development criteria is more restrictive. Obviously, CE may 
lead in some cases to binary component merger and formation of a massive single stars. 
Note however, that even in our standard scenario dominant formation channel,
the first RLOF is stable TTMT from HG star (compare dominant path 3 for model M380.B with
dominant path 1 for model M480.B in Tab. \ref{tab:formation_channels}). 
The second stable RLOF phase would be replaced by CE in the standard model. 
The CE during second RLOF is the main phase which leads the orbit size 
decrease for final BH-BH system to merge within Hubble time 
(see path 3 in Tab. \ref{tab:formation_channels} for model M380.B).

For Model M481.B (revised CE development criteria and modified condition for switch from TTMT 
to nuclear timescale MT) the main BH-BH formation channel also differs from 
the one in the standard Startrack approach (M380.B). Most of the evolution
resembles the scenario described for M480.B: first TTMT initiated by primary, 
then the first BH formation with possible natal kick and the second TTMT initiated by
secondary. However, in model M481.B due to modified condition for switch from
the TTMT to nuclear-timescale stable MT, and the "safety" condition 
(read $\S$ 3.2), when TTMT ends system goes through CE phase. If the system
survives CE, the separation is reduced and the secondary's envelope is ejected 
leaving a naked helium core. Soon thereafter CE (in some cases also already during 
CE), the secondary's core collapse and the second (typically less massive) BH is formed.

   \begin{table*} 
      \caption[]{Local ($z \sim 0$) merger rate densities 
      $\left[\mbox{Gpc}^{- 3} \mbox{yr}^{- 1} \right]$ for different BH-BH formation 
      scenarios. The main formation scenario for given model is marked with the bold text. }
        \label{tab:formation_channels}
     $$ 
         \begin{array}{llccc}
            \hline
            \noalign{\smallskip}
            \mbox{No.} & \mbox{BH-BH formation scenario}  & \cal{R}_{\mbox{M480.B}}  & \cal{R}_{\mbox{M481.B}} & \cal{R}_{\mbox{M380.B}} \\
            \noalign{\smallskip}
            \hline
            \hline
            \noalign{\smallskip}
            \mbox{1.}&\mbox{MT1(1/2/4/5-1/4)/MT2(7-4) BH1 MT2(14-2/4/5/7/8) BH2} & \mbox{\bf{83.1}} & \mbox{0.4} & \mbox{0.3}\\
            \mbox{2.}&\mbox{MT1(1/2/4-1) BH1 MT2(14-2/4) SW CE2(14-4/5:14-7/8/14) BH2} & \mbox{0.4} & \mbox{\bf{12.5}} & \mbox{0.0} \\
            \mbox{3.}&\mbox{MT1(1/2/4-1) BH1 CE2(14-4/5:14-7/8/14) BH2} &  \mbox{2.1} & \mbox{2.0} & \mbox{\bf{51.2}} \\
            \mbox{4.}&\mbox{MT1(1/2/4-1) BH1 MT2(14-2/4) CE2(14-4/5:14-7/8/14) BH2} &  \mbox{1.1} & \mbox{1.0} & \mbox{1.6}\\
            \mbox{5.}&\mbox{MT1(4-2/4) MT2(4/5/7/8-4) CE12(4/5/8-4:14-7) BH1 BH2} & \mbox{1.4} & \mbox{1.4} & \mbox{2.0} \\
            \mbox{6.}&\mbox{Other scenarios} & \mbox{0.3} & \mbox{0.6} & \mbox{6.6} \\
            \hspace{0.5cm}
            \mbox{}&\mbox{Total} & \mbox{88.4} & \mbox{17.9} & \mbox{61.7} \\
            \hline
         \end{array}
     $$ 
     MT1-stable RLOF, donor is initially more massive star;\\ 
     MT2-stable RLOF, donor is initially less massive star; \\
     BH1-formation of black hole by initially more massive star; \\
     BH2-formation of black hole by initially less massive star; \\
     CE1-common envelope initiated by initially more massive star; \\
     CE2-common envelope initiated by initially less massive star; \\
     CE12-double common envelope initiated by two giants; \\
     SW - switch from TTMT to CE based on the revised condition (see \S 3.2 and \S 5). \\

The numeric types are consistent with \cite{Hurley_2002}: \\
0 – main sequence star with M<=0.7 Msun (deeply or fully convective) \\
1 – main sequence star with M>0.7 Msun \\
2 – Hertzsprung gap star \\
3 – first giant branch star \\
4 – core helium burning star \\
5 – early asymptotic giant branch star \\
6 – thermally pulsing asymptotic giant branch star \\
7 – main sequence naked helium star \\
8 – Hertzsprung gap naked helium star \\
9 – giant branch naked helium star \\
10 – helium white dwarf \\
11 – carbon or oxygen white dwarf  \\
12 – oxygen or neon white dwarf \\
13 – neutron star \\
14 – black hole \\
15 – massless remnant \\

The values in the brackets, e.g., 1/2/4 refer to different possible variants of the evolutionary 
types of binary components during given phase. The type of initially more massive component is given
as the first and the type of initially less massive component as a second (after dash,-). In the case
of CE phase type at the onset are given as the first, next types after envelope ejection 
(after colon).
     
   \end{table*}

\subsection{BH-NS formation scenarios} \label{AppendixB}
\begin{figure*} 
   \centering
   \includegraphics[width=180mm]{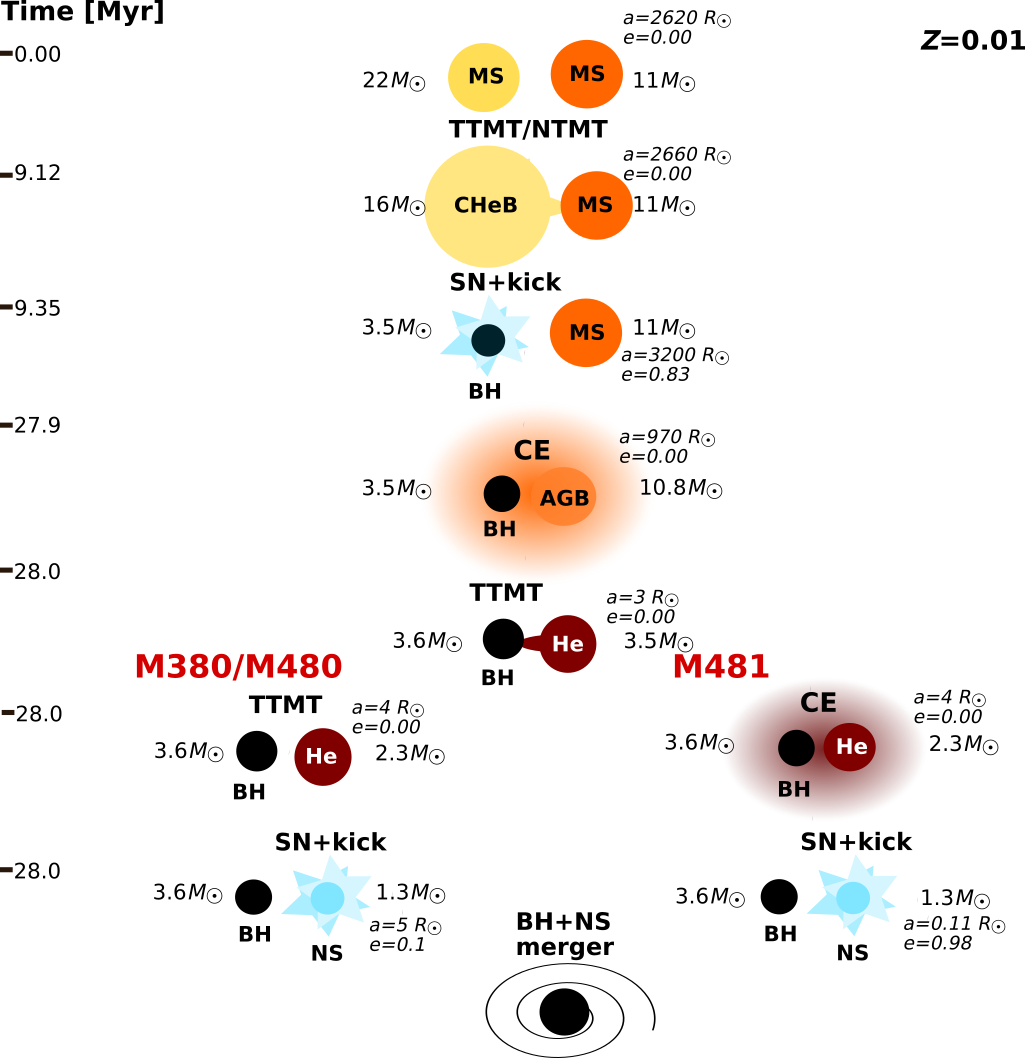}
   \caption{We present the typical local ($z \approx 0$) BH-NS mergers formation scenarios 
   for three tested models M380, M480 and M481 (for models M380 and M480 the scenario is the same). 
   For all models we begin with the same initial conditions: a wide binary system with 22$\msun$ 
   primary and 11$\msun$ secondary on the orbit of $\sim 2600\rsun$ with metallicity $Z=0.01$. 
   Most of the evolutionary phases are common for all three models. 
   When the primary leaves main sequence 
   it expands, first as HG star and then as core helium burning giant. The system goes through 
   stable RLOF phase with TTMT and nuclear-timescale MT episodes. After MT the donor looses its 
   envelope and soon explodes as SN. A low mass BH (3.5$\msun$) is formed with the 
   accompanying high natal kick which causes the system eccentricity increase to e=0.83. Next, the 
   secondary leaves main sequence and begins to expands. At the same time the orbit gets circulized 
   by the tides. System goes thorough the CE phase, the secondary looses its envelope. After CE 
   separation decreases to only 3$\rsun$. Close system of BH and helium star is formed. 
   As after CE the secondary is a main sequence or HG naked helium star, it expands initiating a 
   stable RLOF. The system goes through next TTMT which in model M481 finishes with the 
   second CE phase and formation of very close (0.11$\rsun$) BH-NS system. In the case of models M380 
   and M480 once TTMT ends, system remains a close BH-He star binary and soon the secondary explodes 
   forming a NS. "NNMT" - nuclear-timescale stable MT; "He"- stripped helium core; 
   "AGB"-asymptotic giant branch.}
   \label{BHNS_diagram}
    \end{figure*}

In Table \ref{tab:formation_channels_BHNS} we provide formation scenarios of local ($z\approx 0$) 
BH-NS mergers for tested models with revised CE development criteria (M480.B and M481.B) and 
standard {\tt Startrack} criteria (M380.B). For each model the dominant formation scenario 
which constitutes the highest merger rate density fraction is marked with the bold text. For 
two models M380.B and M480.B the dominant formation scenario is the same, however, for model 
M480.B it constitutes over 70\%  while for M380.B about 50\% of total BH-NS merger rates. The evolutionary diagrams with the dominant formation scenarios are shown in Figure \ref{BHNS_diagram}.

Most of the evolutionary phases are common for all three tested models. Evolution scenario goes 
as follows: once primary, initially more massive star, leaves main sequence, it expands and
initiates first stable RLOF. System goes through the stable MT, after which the donor (primary) star 
looses its envelope becoming a stripped helium star. Next, primary forms a BH through 
SN explosion or via direct collapse (depends on the final star mass). 
When the secondary star leaves its main sequence it expands and initiates unstable MT. 
The CE development leads to the secondary's envelope removal and significant contraction 
of the system separation. Close system of BH and naked helium star is formed.
As after CE phase the secondary is a main sequence or HG helium star, it still expands
initiating a stable RLOF. At the end of the TTMT, the evolutionary scenarios for 
different models begin to diverge. 
In the case of models M380.B and M480.B, after TTMT system is close BH-He binary. Next, there is a SN explosion and the NS formation. In the scenario M481.B, due to 
modified condition for the switch between the TTMT and nuclear-timescale stable MT, and the "safety" 
condition (read \S \ref{timescales}), the system ends stable RLOF with CE phase.
The orbit is tightened once again. During the second CE phase the NS is formed.

   \begin{table*} 
      \caption[]{Local ($z \sim 0$) merger rate densities 
      $\left[\mbox{Gpc}^{- 3} \mbox{yr}^{- 1} \right]$ for different BH-NS formation 
      scenarios. The main formation scenario for given model is marked with the bold text. }
        \label{tab:formation_channels_BHNS}
     $$ 
         \begin{array}{llccc}
            \hline
            \noalign{\smallskip}
            \mbox{No.} & \mbox{BH-NS formation scenario}  & \cal{R}_{\mbox{M480.B}}  & \cal{R}_{\mbox{M481.B}} & \cal{R}_{\mbox{M380.B}} \\
            \noalign{\smallskip}
            \hline
            \hline
            \noalign{\smallskip}
            \mbox{1.}&\mbox{MT1(1/2/4-1) BH CE2(14-4/5:14-7/8/14) MT2(14-8/9)  NS} &  \mbox{\bf{11.6}}  & \mbox{0.7}& \mbox{\bf{6.6}}\\
            \mbox{2.}&\mbox{MT1(1/2/4/5-1/4)/MT2(7-4) BH MT2(14-2/4/5/7/8) NS} & \mbox{{1.2}} & \mbox{0.0} & \mbox{0.0}\\
            \mbox{3.}&\mbox{MT1(1/2/4-1) BH CE2(14-4/5:14-7/8/14) NS} &  \mbox{1.0} & \mbox{0.8} & \mbox{3.5} \\
            \mbox{4.}&\mbox{MT1(1/2/4-1) BH CE2(14-4/5:14-7/8) MT2(14-8) SW CE2(14-9:14-13) NS} &  \mbox{0.0} & \mbox{\bf{1.5}} & \mbox{0.0}\\
            \mbox{5.}&\mbox{Other scenarios} & \mbox{1.8} & \mbox{1.1} & \mbox{3.0} \\
            \hspace{0.5cm}
            \mbox{}&\mbox{Total} & \mbox{15.6} & \mbox{4.1} & \mbox{13.1} \\
            
            \hline
         \end{array}
     $$ 
     Abrrevations are the analogous with the described in Table \ref{tab:formation_channels}.
\end{table*}

\label{sec:appendix}

\section{Conclusions and discussion}

In this work, we implement into {\tt StarTrack} population synthesis code the revised RLOF stability
diagram for binary systems with initially massive donors ($M_{\rm{ZAMS}}>18 \msun$). The revised
diagram is based on MT simulations performed by \cite{2017MNRAS.465.2092P} using 1D hydrodynamical
stellar code {\tt MESA}.  They found that RLOF is stable over a much wider parameter space than
previously thought if the MT is allowed to proceed until outflows from the outer Langrangian points
develop. For one tested model, we also modify conditions for switching between TTMT and nuclear
timescale MT. We used the most updated version of {\tt StarTrack} code to calculate local merger 
rate density and mass ratio distribution of the synthetic population of DCOs mergers.

We compare the results of the revised RLOF treatment with the standard {\tt StarTrack}. 
We also check if new results are consistent with the recent estimates of LIGO/Virgo 
collaboration, comparing local merger rate density, the fraction of unequal mass mergers 
and the shape of BH-BH mergers mass distribution.

We present the results of three models; our standard {\tt StarTrack} input physics model with delayed 
SN engine (M380.B) and two models with revised RLOF treatment (M480.B and M481.B). 
Model M480.B differs from our standard M380.B model only by the revised CE development 
criteria while M481.B additionally includes the modified condition for the switch from 
TTMT to nuclear timescale MT.  

Various changes related to assumptions about RLOF physics may strongly affect 
(sometimes in a non-intuitive way) the formation of DCO mergers: impacting merger 
rates and physical properties of merging objects. For the three presented models
not only different local merger rate densities or BH mergers 
mass distributions were obtained but also the dominant evolutionary scenarios for 
BH-BH formation changes. 

In model M480.B (with revised CE development criteria) the most common scenario 
for BH-BH merger formation is without any CE phase. In model M481.B
(with revised CE development criteria and modified switch from TTMT to nuclear 
timescale MT) BH-BH dominant formation scenario involves CE that is preceded by short TTMT
phase.

Among the tested models, we did not find one that would be fully consistent with the
merger rates for NS-NS, BH-NS, and BH-BH given by LIGO/Virgo and at the same time 
would produce the reported more massive BH mass distribution for BH-BH mergers. 
While the shape of the more massive BH mass 
distribution in model M480.B fits well two exponent values and the 
break point given \citep{catalog03}, the total rates for BH-BH mergers are too high 
by the factor of $\sim 2-3$. Additionally, the fraction of unequal BH-BH mergers 
is not high enough. For Model M481.B the total local rate density for all types of DCO 
mergers are well consistent with the most recent LIGO/Virgo estimates, but the shape of the 
more massive BH distribution seems too steep. Obviously, we could try to modify RLOF 
physics further or even change other parts of input physics (e.g., increase BH natal 
kicks to lower BH-BH merger rates) to fit LIGO/Virgo data for some models.
However, the point of this paper was only to demonstrate how uncertain ingredients of 
population synthesis codes may dramatically alter predictions. 

We checked that the fraction of accreted mass during non-conservative RLOF (non-degenerate 
accretors, see \S \ref{sec: method}) does not significantly influence the population of 
BH-BH and BH-NS mergers in tested model M380.B. Beside standard {\tt Startrack} value 
$f_{\rm a}=0.5$ , we tested models with values $f_{\rm a}=0.8$ and $f_{\rm a}=0.2$ for 
accreted mass fraction (the fraction $1-f_{\rm a}$ of transferred mass is ejected from 
the system). We found that merger rates for BH-BH and BH-NS systems change about 
$3 \%$ for $f_{\rm a}=0.8$ and about $20 \%$ for $f_{\rm a}=0.2$ comparing to the model 
with $f_{\rm a}=0.5$. Also mass and mass ratio distribution for three tested values are similar, 
with $m_1$ (more massive of merging components) slightly shifted to the lower mass values 
in the model with the lowest accretion fraction $f_a=0.2$. Different adopted values of 
$f_a$ influence, however, the local merger rate density for NS-NS systems. In both 
$f_a=0.2$ and $f_a=0.8$ models the rate for NS-NS systems decreased by a factor of 
$\sim 2$ comparing to the standard $f_a=0.5$ model.
The drop in NS-NS rates is a complex mix of multiple overlapping physical processes, e.g., 
different amount of lost mass (together with orbital angular momentum) leads to 
different system separation after the first RLOF, which further leads to a different 
evolutionary type of the donor during the second RLOF (CE phase with HG donor ends 
with a system merger).

In previous works based on {\tt StarTrack} code, the two main variants of CE treatment 
were usually considered, referred to as submodel A and B. 
Submodel B (the one presented in this work, see \S \ref{sec: method}) assumes that all 
HG donor systems merge during the CE phase. 
In contrast, submodel A let such binary systems survive if the energy budget calculations 
allow. We decided to present only the results for the submodel B for two main reasons. 
First, we checked that typically HG donors which initiate CE in {\tt StarTrack} 
are only partially expanded stars, shortly after their main sequence. Such stars likely 
do not have a clear core-envelope boundary, and therefore, it seems more physical to assume 
their merger (as we do for main sequence stars). Second, the merger rates for DCOs, especially 
BH-BH systems in submodel A (see \cite{2020A&A...636A.104B}) are often much larger 
(up to 30 times) than estimated by LIGO/Virgo. 

Our rapid population synthesis study's central message is that RLOF treatment's choice 
produces noticeably different populations of GW sources. 
Our results highlight the need for caution in information inference from stellar and 
binary evolution models as applied to LIGO/Virgo results. Similar cautionary notes 
are also found through more detailed evolutionary calculations with hydrodynamical 
stellar codes \citep{Klencki2020,decin2020evolution}. 
Suppose a population synthesis study is performed with variations of only natal 
kicks and CE efficiency, and the match to LIGO/Virgo rates is achieved for some 
specific CE efficiency and natal kicks assumption. In that case, it cannot be 
expected to mean that we have already constrained these uncertain parts of input physics.

\begin{acknowledgements}
We  would  like  to  thank  anonymous  referee  for their useful comments.
K.B. and A.O. acknowledge support from the Polish National Science Center (NCN) grant
Maestro (2018/30/A/ST9/00050). 
N.I. acknowledges support from CRC program and
funding from NSERC Discovery under Grant No. NSERC RGPIN-2019-04277.
\end{acknowledgements}

\bibliographystyle{aa}
\bibliography{ms}

\clearpage

%
%

\end{document}